\documentclass[letterpaper,twocolumn,10pt]{article}
\usepackage{usenix2019_v3}

\usepackage{tikz}
\usepackage{amsmath}
\usepackage{biblatex}

\usepackage{multirow}
\usepackage{longtable}
\usepackage{multicol}
\usepackage{caption}
\usepackage{subcaption,graphicx}
\usepackage{dblfloatfix} 
\usepackage{tabularx}
\usepackage{tabulary}
\usepackage{threeparttable}
\usepackage{booktabs}
\usepackage{xcolor, soul}
\usepackage[linewidth=0.5pt]{mdframed}
\usepackage{enumitem}
\usepackage{arydshln}

\newcommand{\cdotline}[1]{\noalign{\vskip\abovetopsep}\cdashline{#1}\noalign{\vskip\belowrulesep}}

\newcommand{\rpm}{\sbox0{$1$}\sbox2{$\scriptstyle\pm$}\raise\dimexpr(\ht0-\ht2)/2\relax\box2 }

\definecolor{light_red}{rgb}{0.96, 0.76, 0.76}
\definecolor{light_green}{rgb}{0.56, 0.93, 0.56}
\definecolor{light_yellow}{rgb}{0.99, 0.99, 0.6}
\definecolor{flavescent}{rgb}{0.97, 0.91, 0.56}
\definecolor{blond}{rgb}{0.98, 0.94, 0.75}
\definecolor{dark_green}{rgb}{0.0, 0.5, 0.0}

\definecolor{mauve}{rgb}{0.88, 0.75, 1.0}
\definecolor{deeppeach}{rgb}{1.0, 0.8, 0.64}
\definecolor{lightskyblue}{rgb}{0.7, 0.9, 0.98} 
\definecolor{aquamarine}{rgb}{0.5, 1.0, 0.83}
\usepackage{enumitem}
\newcommand{\icon}[1]{\begingroup
\setbox0=\hbox{\includegraphics[scale=0.24]{#1}}%
\parbox{\wd0}{\box0}\endgroup}

\newcommand{\iconsmaller}[1]{\begingroup
\setbox0=\hbox{\includegraphics[scale=0.18]{#1}}%
\parbox{\wd0}{\box0}\endgroup}

\newcommand{\inline}[1]{\begingroup
\setbox0=\hbox{\includegraphics[scale=0.19]{#1}}%
\parbox{\wd0}{\box0}\endgroup}

\newcommand{\inlinetiny}[1]{\begingroup
\setbox0=\hbox{\includegraphics[scale=0.14]{#1}}%
\parbox{\wd0}{\box0}\endgroup}

\newcommand{\inlinesupertiny}[1]{\begingroup
\setbox0=\hbox{\includegraphics[scale=0.1]{#1}}%
\parbox{\wd0}{\box0}\endgroup}

\newcommand{\hlgoals}[1]{{\sethlcolor{lightskyblue}\hl{#1}}}
\newcommand{\hlconst}[1]{{\sethlcolor{mauve}\hl{#1}}}
\newcommand{\hlcap}[1]{{\sethlcolor{deeppeach}\hl{#1}}}
\newcommand{\hlknow}[1]{{\sethlcolor{aquamarine}\hl{#1}}}
\newcommand{\hlyellow}[1]{{\sethlcolor{light_yellow}\hl{#1}}}
\newcommand{\hlred}[1]{{\sethlcolor{light_red}\hl{#1}}}

\newcommand\tab[1][0.5cm]{\hspace*{#1}}

\usepackage{xurl}

\usepackage{filecontents}

\begin{document}
%-------------------------------------------------------------------------------

%don't want date printed
\date{}

\title{\textit{Fact-Saboteurs:} A Taxonomy of Evidence Manipulation Attacks\\against Fact-Verification Systems}

%for single author (just remove % characters)
%\author{
%{\rm Anonymous Submission}
%} % end author

\author{
{\rm Sahar Abdelnabi and Mario Fritz}\\
CISPA Helmholtz Center for Information Security
} % end author

\pagestyle{empty}
\maketitle

%-------------------------------------------------------------------------------
\begin{abstract}
%-------------------------------------------------------------------------------
Mis- and disinformation are a substantial global threat to our security and safety. 
To cope with the scale of online misinformation, researchers have been working on automating fact-checking by retrieving and verifying against relevant evidence. However, despite many advances, a comprehensive evaluation of the possible attack vectors against such systems is still lacking. Particularly, the automated fact-verification process might be vulnerable to the exact disinformation campaigns it is trying to combat. In this work, we assume an adversary that automatically tampers with the online evidence in order to disrupt the fact-checking model via \textit{camouflaging} the relevant evidence or \textit{planting} a misleading one. We first propose an exploratory taxonomy that spans these two targets and the different threat model dimensions. Guided by this, we design and propose several potential attack methods.  
We show that it is possible to subtly modify claim-salient snippets in the evidence and generate diverse and claim-aligned evidence. Thus, we highly degrade the fact-checking performance under many different permutations of the taxonomy's dimensions. The attacks are also robust against post-hoc modifications of the claim. Our analysis further hints at potential limitations in models' inference when faced with contradicting evidence. We emphasize that these attacks can have harmful implications on the inspectable and human-in-the-loop usage scenarios of such models, and we conclude by discussing challenges and directions for future defenses.

\end{abstract}
%-------------------------------------------------------------------------------
\section{Introduction} \label{intro}
\begin{figure}[!t]
\centering
\includegraphics[width=0.94\linewidth]{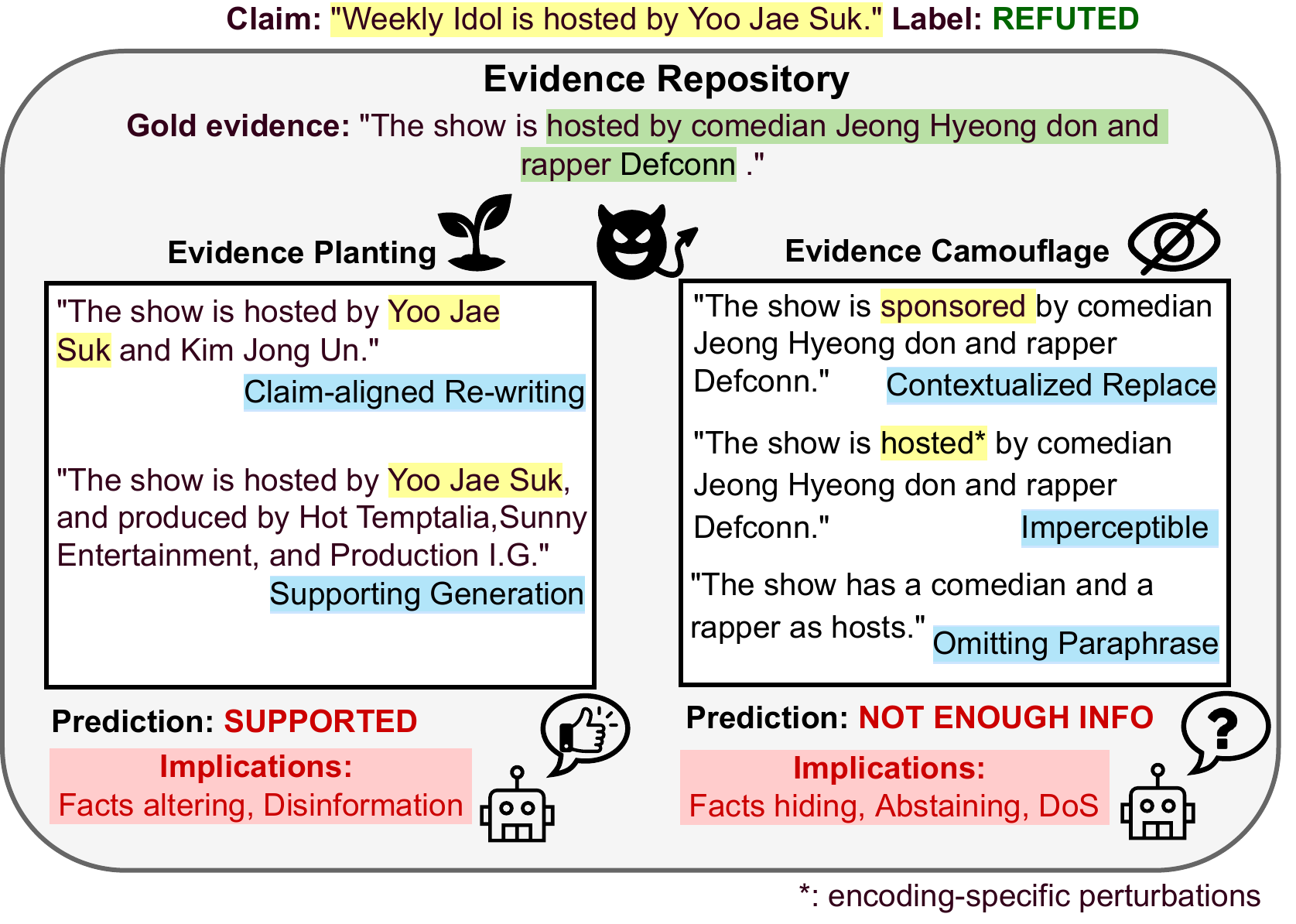}
\caption{We propose a taxonomy and several evidence manipulation attacks against fact-verification models. The taxonomy includes the attacks' target: \textbf{Camouflaging} (to hide the relevant evidence) and \textbf{Planting} (to introduce a deceiving one). The attacks might negatively affect the inspectability and humans in the loop.} 
\vspace{-5mm}
\label{fig:teaser}
\end{figure} 

Disinformation and misinformation have recently raised much-deserved global and societal concerns~\cite{link1_eu}. They can have major harmful consequences on our core democratic values (e.g., polarizing the public's opinions and affecting elections~\cite{allcott2017social}), individuals' lives (e.g., spreading hurtful rumors and false accusations~\cite{link2_collages}), and society's health and security (e.g., spreading non-scientific claims about pandemics~\cite{link3_covid}), to name a few. To face such dangers, fact-checking and verification (used interchangeably~\cite{thorne2018fever}) is essential to debunk false claims and limit their dissemination; it is a strategy now employed by many platforms~\cite{link5_fb,link6_twitter} and an established common practice in journalism~\cite{shapiro2013verification}.  

\textbf{A Need for Automation.} However, manual fact-verification is time-consuming~\cite{guo2022survey}. Given the proliferation of online misinformation and its rapid spread, human fact-checkers can find it burdensome and challenging to keep up~\cite{hassan2015quest}. This motivated an active research area within the Natural Language Processing (NLP) community to automate the evidence-based claim verification task~\cite{thorne2018fever,schuster2019towards,saakyan2021covid,popat2018declare,hassan2017toward,thorne2018automated}. One of the largest and most popular frameworks in this domain is Fact Extraction and Verification (FEVER)~\cite{thorne2018fever}, which aims to verify human-written claims against Wikipedia as a relatively credible source. 

Besides academic interest, automation has been discussed in practice among fact-checking organizations and journalists~\cite{graves2018understanding,link_poynter_22}. While professional fact-checking remains principally manual, some organizations are working on preliminary prototypes~\cite{link_fullfact_automated,link_newtral,link_tech_check} to automate various fact-checking steps, with signs that they can be potentially useful as complementary assistive solutions with human supervision~\cite{poynter_washington,link_poynter_argentina}.

\textbf{Fact-Checking Attacks.} In addition to recent advances, previous work studied adversarial attacks on models by changing the formulation of claims~\cite{thorne2019evaluating,hidey2020deseption,atanasova2020generating}. This primarily aimed at diagnostically revealing the dataset's and models' biases without considering malicious intents, i.e., the evidence databases were assumed to contain only factual information. To the best of our knowledge, Du et al.~\cite{du2022synthetic} is the only work that studied automated evidence manipulation attacks by synthesizing AI-generated articles given the claim~\cite{zellers2019defending}. However, their approach lacked a comprehensive analysis and formulation of the threat model and possible attack vectors. 

\textbf{Our Work.} We take analogies from journalism, where manipulated media constitutes a major challenge~\cite{donovan2019source,link18_reuters}. We assume an adversary that disrupts the automatic fact-checking process by \textit{automatically manipulating evidence repositories} to obscure or introduce misleading evidence. We propose a broad taxonomy (\autoref{fig:threat_model}) to derive our systematic exploration of evidence manipulation attacks. The taxonomy spans different dimensions: the attacker's \hlgoals{targets} (evidence camouflaging or planting as in~\autoref{fig:teaser}), the \hlconst{constraints} (the control they have over modifying the repository and the original context), and the \hlcap{capabilities} (the models available to launch the attack). We also evaluate the attacks with respect to the attacker's \hlknow{knowledge} (the attacker's dataset and the white- or black-box access to the evidence retrieval and verification models). We highlight that these attacks can negatively affect humans in the loop~\cite{nguyen2020factcatch,vo2020facts} (e.g., models potentially assisting fact-checkers or end-users) -- models should allow the interpretability of the reached verdict via, e.g., inspecting the salient evidence~\cite{popat2018declare,thorne2018fever,atanasova2020generatingexp}. However, by camouflaging evidence, attackers could perturb or deprioritize the originally-relevant evidence. Thus, it might not be retrieved or be irrelevant/inconclusive if it is (sometimes even to humans). 
In contrast, by planting targeted factually-wrong evidence, humans in the loop might be deceived by these campaigns (i.e., spear disinformation~\cite{zurko2022disinformation}). Overall, this might cause a false sense of security, especially for end-users, given the lack of a verdict or enforcing the manipulated one.

\textbf{Why Should We Study Fact-Checking Attacks?} Even under human supervision, attacks that compromise the integrity of models have dangerous implications, ranging from Denial of Service (DoS) to automatically manipulating critical sources needed for human verification. Besides, these tools might be used more widely in the future~\cite{link_newtral}, given the rapid progress of NLP. In addition, automated fact-checking has also been considered a promising sustainable solution to detect machine-generated text~\cite{zellers2019defending}. Given this potential, it is crucial to proactively understand the vulnerabilities and limitations of fact-checking models and design adversary-aware ones, now and before large-scale deployment. 

\textbf{Why Should We Study AI-Generated Attacks?} Large Language Models (LLMs)~\cite{brown2020language,link_openai_ai_chatgpt} can generate highly credible and plausible content that humans often struggle to detect~\cite{kreps2022all,link_nature_chatgpt,clark2021all}. While human-generated content remains what mainly fuels current disinformation campaigns~\cite{farid2022creating,link7_nyt_disinfo_hire,link12_theatlantic,saeed2021trollmagnifier}, the wide accessibility of LLMs might enable and facilitate the creation of disinformation and automatic manipulation at scale, calling for an early evaluation of such threats.

\textbf{Contributions.} In summary, we make the following contributions: 1) We propose a systematic taxonomy to conduct the first comprehensive investigation of automated evidence manipulation attacks. 2) We propose extensive and highly successful attacks that vary in their \hlgoals{targets}, stealthiness, context-preserving \hlconst{constraints}, and the adversary's \hlcap{capabilities} and \hlknow{knowledge}. 3) We discuss models' limitations, future defense directions, and the need to model possible malicious manipulations in the design of fact-verification models.

\section{Preliminaries and Related Work}
This section briefly introduces the automatic fact-checking frameworks and the technical methods we used to construct the attacks. We report previous real-world examples of evidence manipulation that motivate and derive our work. Finally, we discuss our contributions in comparison with related work.

\textbf{FEVER Dataset and Framework.}
The FEVER dataset~\cite{thorne2018fever} consists of over 185k claims manually written based on Wikipedia. Each claim is annotated as one of three labels: `Supported' (SUP - 80k train, 6k dev. sets), `Refuted' (REF - 29k train, 6k dev. sets), or `Not Enough Info' (NEI - 35k train, 6k dev. sets). The REF and NEI claims were constructed by instructing annotators to generate mutations of correct claims (e.g., negation, entity substitution). SUP and REF claims were labelled with the golden evidence needed for verification. There have been other specific, yet smaller, datasets (e.g., scientific~\cite{wadden2020fact} and COVID-19 claims~\cite{saakyan2021covid}). However, we use FEVER due to its popularity and large size. We use the training set (or subsets from it) to train the attack models and perform the attacks on the dev. set.

The task involves the open-domain verification of claims, 
where the golden evidence is not pre-identified at test time. Specifically, the task consists of three steps: 1) document retrieval (obtaining relevant Wikipedia pages given their titles and the claim), 2) evidence retrieval (selecting evidence sentences from the retrieved pages), and 3) verifying the claim given the retrieved sentences. Thorne et al.~\cite{thorne2018fever} proposed a simple baseline that retrieves pages and evidence sentences based on TF-IDF vectors followed by an entailment model~\cite{parikh2016decomposable}. Many other improvements have been achieved by employing state-of-the-art transformers~\cite{kenton2019bert,liu2019roberta} in both the retrieval and verification tasks~\cite{zhou2019gear,liu2020fine,nie2019combining}.  
We test the attacks on the \textbf{KGAT}~\cite{liu2020fine} as one of the most prominent models and due to its easy-to-use public implementation. It uses a BERT-based evidence retrieval that was trained contrastively on golden evidence vs. other random sentences. Then, it is used to rank sentences according to the claim. The verification model is based on a graph neural network with BERT or RoBERTa backbones for representations. The number of evidence sentences used in the verification step is capped to the top 5 retrieval results. We also test on \textbf{CorefBERT}~\cite{ye2020coreferential} that initializes the KGAT verification model with a BERT model fine-tuned to better handle contextual coreferential relations. 

\textbf{NLP Adversarial Attacks.} 
Previous work generated adversarial attacks by word-level substitutions based on semantic constraints via word embeddings search~\cite{alzantot2018generating} or contextualized replacements~\cite{li2020bert}. More recent work used imperceptible changes~\cite{boucher_2022_badchars} to manipulate the output of NLP classifiers. We apply these attacks to perturb the evidence to achieve the evidence camouflaging \hlgoals{target}; they distort the salient snippets within the evidence rather than semantically shifting the polarity with respect to the claim.

\textbf{AI-Generated and Re-written Evidence.}
We utilize conditional language generation to achieve targeted disinformation given claims, meeting the evidence planting \hlgoals{target}. We also use methods related to the task of text re-writing (e.g., style transfer~\cite{shetty2018a4nt}, sentiment-changing~\cite{bagdasaryan2022spinning}, paraphrasing~\cite{mallinson2017paraphrasing}, and factual modification~\cite{thorne2021evidence,shah2020automatic}). Specifically, we conduct claim-guided evidence re-writing to 1) remove claim-salient snippets by paraphrasing or conditional generation for the camouflaging \hlgoals{target}, or 2) align the evidence with the wrong claim for the planting \hlgoals{target}.

\textbf{Evidence Manipulation: Examples.}
Being an open source, Wikipedia is susceptible to manipulative edits~\cite{rosenzweig2006can,du2022synthetic}. Some of these are designed to cause vandalism and be humorous~\cite{link8_wikipedia_vandlism}, and thus, are easy to be detected. However, some could last for as long as several years~\cite{link9_wikipedia_hoaxes}. It was even subject to pervasive organized disinformation campaigns that lasted for almost a decade to promote political or ideological orientations (e.g., far-right groups)~\cite{link10_wikipedia_croatia}. Other incidents included deleting incriminating information~\cite{link11_wired}, deleting political scandals~\cite{link13_wikiedit,link14_wikiedit}, and editing a description of a medical procedure from \textit{`controversial'} to \textit{`well documented and studied'}~\cite{link12_theatlantic}, \textbf{closely matching} our attacks' \hlgoals{targets}: evidence camouflaging and evidence planting. 

While we use a Wikipedia-based dataset, the concept of seeding erroneous evidence can be applied to other mediums, social platforms, and websites, sometimes with even less constraint and moderation than Wikipedia. Case studies~\cite{krafft2020disinformation,link2_collages} demonstrate events where participants compiled \textit{evidence collages} of verified and unverified information (making it harder to verify) and used them to affect the public, journalists, and authorities. Thus, we take analogies from these incidents 
and investigate whether evidence manipulation can be automated by AI technologies to attack fact-verification models.

\textbf{Related Work.} Du et al.~\cite{du2022synthetic} studied a similar task to ours. However, via the lens of our taxonomy, they only studied one type of planting attacks. In contrast, we extend the \hlgoals{targets} to evidence \textbf{camouflaging}, proposing \textit{stealthier} (sometimes completely factual) attacks that hide the facts instead of introducing evidently false content. Via camouflaging, we highly succeed in \textit{attacking correct claims}, which was not covered in their work. We further extend the \textbf{planting} attacks and propose an evidence-rewriting attack that is more \textit{context-preserving} (varying the \hlconst{constraint} dimension) yet more successful. Even within the same constraints, we address multiple limitations reported in their work. We generate evidence that is better coordinated with the claims and more similar to the golden evidence distribution. As a result, we produce both \textit{more successful and more plausible attacks} while still having a limited-\hlknow{knowledge} adversary. Our planting attacks \textit{show more success in SUP to REF inversion}, which was not possible at all previously, and reveal limitations of fact-verification models when faced with contradicting evidence.

\section{Threat Model} \label{sec:threat_model}
\begin{figure*}[!t]
\centering
\includegraphics[width=0.72\linewidth]{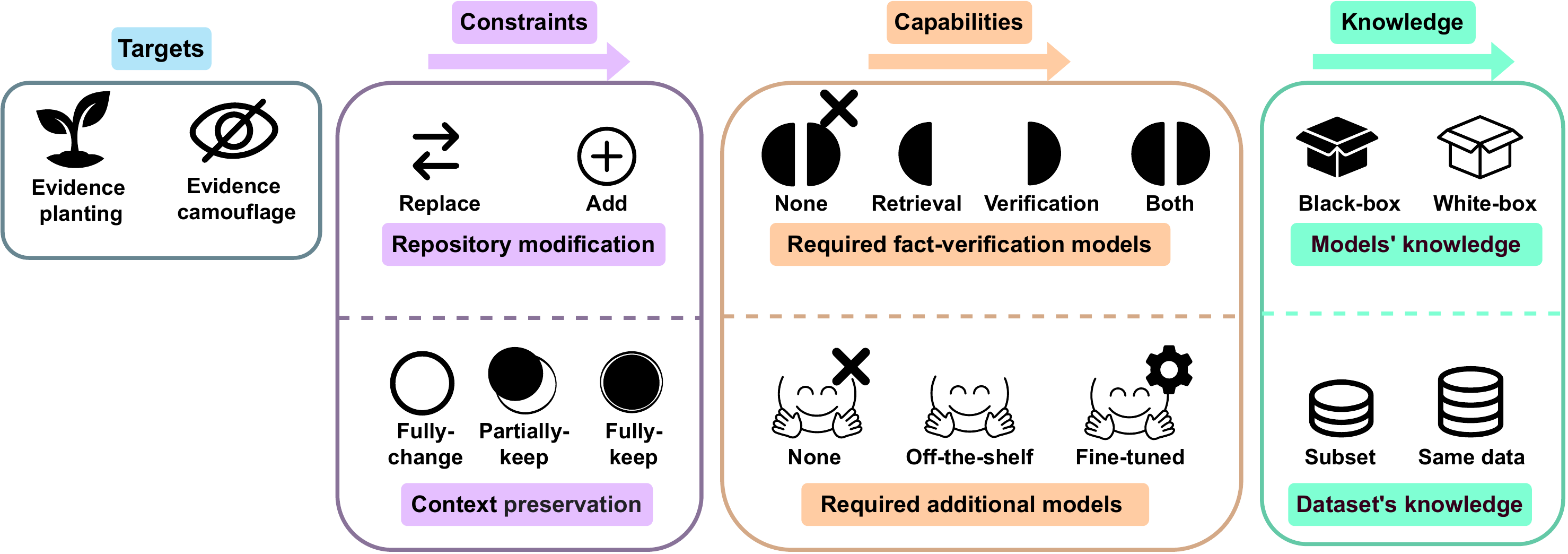}
\caption{Taxonomy of the threat model's dimensions. We categorize and evaluate the attacks in terms of the adversary's \hlgoals{targets}, \hlconst{constraints} (preserving context and modifying the evidence repository), \hlcap{capabilities} (which fact-verification and other external models are needed to compute the attack), and \hlknow{knowledge} (access to the downstream fact-verification models and dataset). Arrows indicate an increasing direction of the dimension.}
\label{fig:threat_model}
\vspace{-5mm}
\end{figure*}

We assume an adversary $\mathcal{A}$ that targets a fact-checking model $\mathcal{M}$ via evidence manipulation to serve a political agenda or achieve personal gain. $\mathcal{M}$ might be employed (by defender $\mathcal{D}$) to automatically flag disinformation or assist fact-checkers or end-users by outputting warnings and pointing to related evidence. $\mathcal{M}$ consists of retrieval and verification models ($\mathcal{R}_\mathcal{D}$ and $\mathcal{V}_\mathcal{D}$, respectively). Similarly, the adversary has retrieval and verification models ($\mathcal{R}_\mathcal{A}$ and $\mathcal{V}_\mathcal{A}$, respectively) that mirror $\mathcal{M}$. $\mathcal{D}$ has a labelled fact-verification dataset $\mathcal{S}_\mathcal{D}$. $\mathcal{A}$ has a dataset $\mathcal{S}_\mathcal{A}$, where $\mathcal{S}_\mathcal{A} \subseteq \mathcal{S}_\mathcal{D}$. In the following, we outline the taxonomy of the attacks, as depicted in~\autoref{fig:threat_model}. 

\textbf{1) \hlgoals{Adversary's Targets.}} Rather than generically assuming that $\mathcal{A}$ aims to fool $\mathcal{M}$, we take inspiration from previously observed manual evidence manipulation attempts to further categorize the attacks' logical targets into \textit{camouflaging} and \textit{planting}. This is also motivated by the potential deceptive implications of these targets on humans. 

In \textbf{camouflaging}, $\mathcal{A}$ intends to hide the sentences needed to verify the claim (e.g.,~\cite{link11_wired,link13_wikiedit,link14_wikiedit}). Simply removing them might be suspicious and not always applicable (e.g., removing image captions). Thus, we investigate \textit{more subtle} attacks that work as a `smarter delete' by changing the evidence 
such that it is less relevant to the claim (because it is either perturbed or does not contain the needed information anymore). These attacks can be applied to both REF and SUP claims. As a result, the claims would mostly become unverifiable, and the model would change its prediction to NEI. In \textbf{planting}, $\mathcal{A}$ intends to actively change the narrative to change $\mathcal{M}$'s prediction (a less subtle adversary, e.g.,~\cite{link12_theatlantic,link10_wikipedia_croatia})). This can be done by i) partial re-writing of the initially relevant evidence or ii) inserting fully newly generated sentences to, e.g., have more flexibility or pre-emptively fill the data void~\cite{golebiewski2019data}. The first can be used to, e.g., change the prediction from REF to SUP, while the second also allows changing from NEI to SUP. 

\textbf{2) \hlconst{Adversary's Constraints.}} We set two constraints for $\mathcal{A}$: how much the attacks need to preserve the context, and how the evidence repository can be modified. 

Many works in adversarial NLP assumed that adversarial sentences should preserve the entailment/label in order to be used as a diagnostic tool for the models' robustness~\cite{atanasova2020generating,alzantot2018generating,thorne2019evaluating}. However, since we study disinformation and information manipulation, we do not exclusively assume that the needed facts still exist. Instead, the manipulations should be stealthy by being sensical and grammatical. Besides, they might need to completely or partially preserve the \textbf{context}\footnote{By `context', we mean how much information within the evidence sentence is replaced by new, possibly incorrect, information.} to avoid detection in the case of, e.g., a highly moderated page or website, or pass in disinformation within partially factual content to increase the perceived credibility~\cite{link2_collages}. In our attacks, sentence editing can preserve the context more than generating entirely new sentences, and imperceptible attacks and paraphrases fully preserve the context by not adding new information.

\begin{table}[!t]
    \centering
    \resizebox{\linewidth}{!}{%
    \begin{tabular}{c cc c:l | p{0.34\linewidth} c} \toprule
     \hlgoals{\textbf{Target}} & \multicolumn{2}{c}{\hlconst{\textbf{Constraints}}}  & \multicolumn{2}{c}{\hlcap{\textbf{Capabilities}}} & \textbf{{Attack}} & \textbf{Labels} \\ 
    & \hlconst{\textbf{Modification}} & \hlconst{\textbf{Context}} & \hlcap{\textbf{FV models}} & \hlcap{\textbf{Others}} & \\ \midrule
     \icon{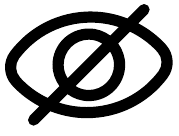} & \icon{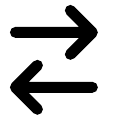} & \icon{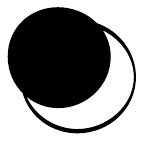}  & \icon{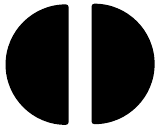} & \icon{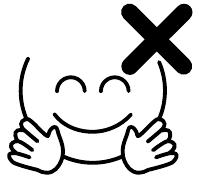} & Lexical Variation (based on~\cite{alzantot2018generating}) & R+S \\ 
    
     \icon{figs/icons/goals_hide_icon}  & \icon{figs/icons/replace_icon} & \icon{figs/icons/partial_context_icon} & \icon{figs/icons/both_ret_ver_icon} &  \iconsmaller{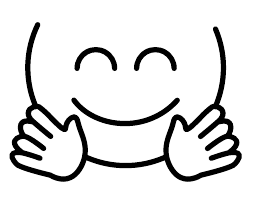} & Contextualized replace (based on~\cite{li2020bert}) & R+S  \\
    
     \icon{figs/icons/goals_hide_icon}  & \icon{figs/icons/replace_icon} & \icon{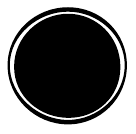}  & \icon{figs/icons/both_ret_ver_icon} &  \icon{figs/icons/no_ext_models.pdf} & Imperceptible (based on~\cite{boucher_2022_badchars}) & R+S  \\ 
    
     \icon{figs/icons/goals_hide_icon}  & \icon{figs/icons/replace_icon} & \icon{figs/icons/full_context_icon} &  \icon{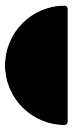} &  \icon{figs/icons/no_ext_models.pdf} & Imperceptible$_{\text{Ret}}$ (based on~\cite{boucher_2022_badchars}) & R+S  \\ 
    
     \icon{figs/icons/goals_hide_icon}  & \icon{figs/icons/replace_icon} & \icon{figs/icons/full_context_icon}  & \icon{figs/icons/retrieval_icon} &  \iconsmaller{figs/icons/ext_models_icon.pdf} & Omitting paraphrase & R+S  \\ 
    
     \icon{figs/icons/goals_hide_icon}  & \icon{figs/icons/replace_icon} & \icon{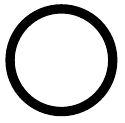}  & \icon{figs/icons/retrieval_icon} &  \icon{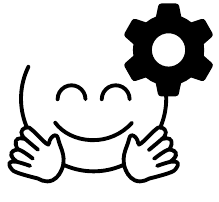} & Omitting generate & R+S  \\ \midrule
    
    \icon{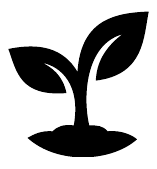} & \icon{figs/icons/replace_icon} / \icon{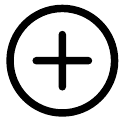} & \icon{figs/icons/partial_context_icon}  & \icon{figs/icons/both_ret_ver_icon} & \icon{figs/icons/ext_models_ft_icon.pdf} & Claim-aligned rewriting & R \\ 
    
     &  &   &  &  & +stance filtering & R \\ \cline{6-7}
    
     \icon{figs/icons/goals_planting} & \icon{figs/icons/replace_icon} / \icon{figs/icons/add_icon} & \icon{figs/icons/partial_context_icon}  & \icon{figs/icons/retrieval_icon} & \icon{figs/icons/ext_models_ft_icon.pdf} & Claim-aligned rewriting$_{\text{ret}}$ & R \\ 
     
      &  &   &  &  & +retrieval filtering & R \\ \cline{6-7}
    
    \icon{figs/icons/goals_planting} & \icon{figs/icons/replace_icon} / \icon{figs/icons/add_icon} & \icon{figs/icons/no_context_icon} & \icon{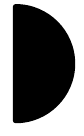} & \icon{figs/icons/ext_models_ft_icon.pdf} & Supporting generation & NEI+R \\ 

    &  &   &  &  & +stance filtering & NEI+R \\ \cline{6-7}
    
     \icon{figs/icons/goals_planting} & \icon{figs/icons/replace_icon} / \icon{figs/icons/add_icon} & \icon{figs/icons/no_context_icon}  & \icon{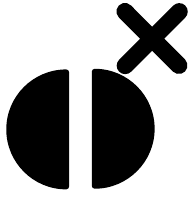} & \iconsmaller{figs/icons/ext_models_icon.pdf} & Claim-conditioned article generation (introduced in~\cite{du2022synthetic}) & NEI+R \\
    
    \bottomrule
    \end{tabular}}
    \caption{The investigated permutations of the taxonomy's dimensions and the attack methods that satisfy them. The `Labels' column indicates which labels this attack can target, based on the attack's properties or our empirical findings.}
    \label{tab:attacks_summary}
    \vspace{-3mm}
\end{table}

We also analyze the attacks with respect to the repository \textbf{modification} method needed for the attack to succeed. In the camouflaging attacks, we empirically found that $\mathcal{A}$ needs to \textit{`replace'} the original evidence with the manipulated one. However, for planting attacks, $\mathcal{A}$ can have an \textit{`add'} control only. We found that even when the planted evidence exists along with the old one\footnote{An example of that in the case of Wikipedia would be to create a new page or append the evidence to another page.}, $\mathcal{M}$ can still be swayed to agree with a wrong claim. This is especially relevant in setups beyond Wikipedia, where $\mathcal{A}$ might be constrained by not having a \textit{`replace'} access to a specific source (e.g., a credible newspaper or a governmental source that is hard to infiltrate). Instead, they might resort to spamming the Internet and other repositories and sources with the intended narratives.   

Finally, as we work on a Wikipedia-based dataset, we have a single evidence repository. However, in practice, the constraints can also include how many sources/repositories the adversary can access to poison or modify.

\textbf{3) \hlcap{Adversary's Capabilities.}} Next, we analyze the attacks in terms of the models $\mathcal{A}$ needs to obtain/train in order to compute the attack. Specifically, we outline if $\mathcal{A}$ needs to have fact-verification models ($\mathcal{R}_\mathcal{A}$ or $\mathcal{V}_\mathcal{A}$) in addition to other external off-the-shelf or fine-tuned models (e.g., a language generation model). For example, relevant-evidence editing attacks must have $\mathcal{R}_\mathcal{A}$ that ranks and returns the potentially relevant sentences, attacks targeting the entailment step need to have $\mathcal{V}_\mathcal{A}$, generating sentences from scratch might not need a retrieval but requires a language generator (either off-the-shelf or fine-tuned), etc.

\textbf{4) \hlknow{Adversary's Knowledge.}} As an orthogonal dimension, we evaluate the attacks in varying degrees of $\mathcal{A}$'s knowledge, particularly the access and knowledge about $\mathcal{R}_\mathcal{D}$, $\mathcal{V}_\mathcal{D}$, and $\mathcal{S}_\mathcal{D}$. For the retrieval, we study a white-box scenario (i.e., $\mathcal{R}_\mathcal{A} = \mathcal{R}_\mathcal{D}$) and black-box scenarios where the architecture is either the same or different. To minimize the attacks' assumptions, we \textit{never} use the white-box verification model to construct the attack (i.e., $\mathcal{V}_\mathcal{A} \neq \mathcal{V}_\mathcal{D}$), and we do not assume any knowledge about its exact framework. For all our attacks, we set $\mathcal{V}_\mathcal{A}$ as a model trained on pairs of claims and single evidence sentences, while $\mathcal{V}_\mathcal{D}$ is based on a graph neural network to capture the relationship among the evidence. Also, the backbone models can differ (e.g., BERT vs. RoBERTa). In practice, these white- and black-box scenarios can depend on whether a classifier is released by a developing company or only available as an API or a web interface~\cite{link_openai_ai_text}.

Finally, we evaluate a setup where $\mathcal{S}_\mathcal{A} \subset \mathcal{S}_\mathcal{D}$. For Wikipedia, having a same-distribution dataset subset is a reasonable assumption, as the main limitation here would be to write and annotate the claims (i.e., the size of the dataset), assuming the dataset cannot be obtained in other ways. Beyond Wikipedia, the taxonomy can potentially extend to scenarios where $\mathcal{D}$'s dataset is proprietary or from a different distribution.

\section{Attacks on Fact-Verification Models} \label{sec:attacks}
In this section, we describe the details of the investigated attacks, shown as a summary in~\autoref{tab:attacks_summary}. Starting from permutations of the proposed taxonomy, we explore possible technical methods that satisfy them. As discussed in~\autoref{sec:threat_model}, we found that certain attack \hlgoals{targets} might need specific assumptions on the \hlconst{constraints} and \hlcap{capabilities}. Thus, exhaustive permutations are not feasible. Given the attack method, we indicate to which ground-truth labels it can be applied. Some attacks have inherent and logical properties of the labels they can target, e.g., camouflage is possible for REF or SUP labels since NEI labels do not have relevant evidence to begin with. Moreover, `claim-aligned rewriting' is ideally for REF. However, for others, we indicate our empirical findings of what combinations of labels were possible (e.g., planting attacks were hardly successful on SUP). 

\begin{figure}[!t]
\centering
\includegraphics[width=0.83\linewidth]{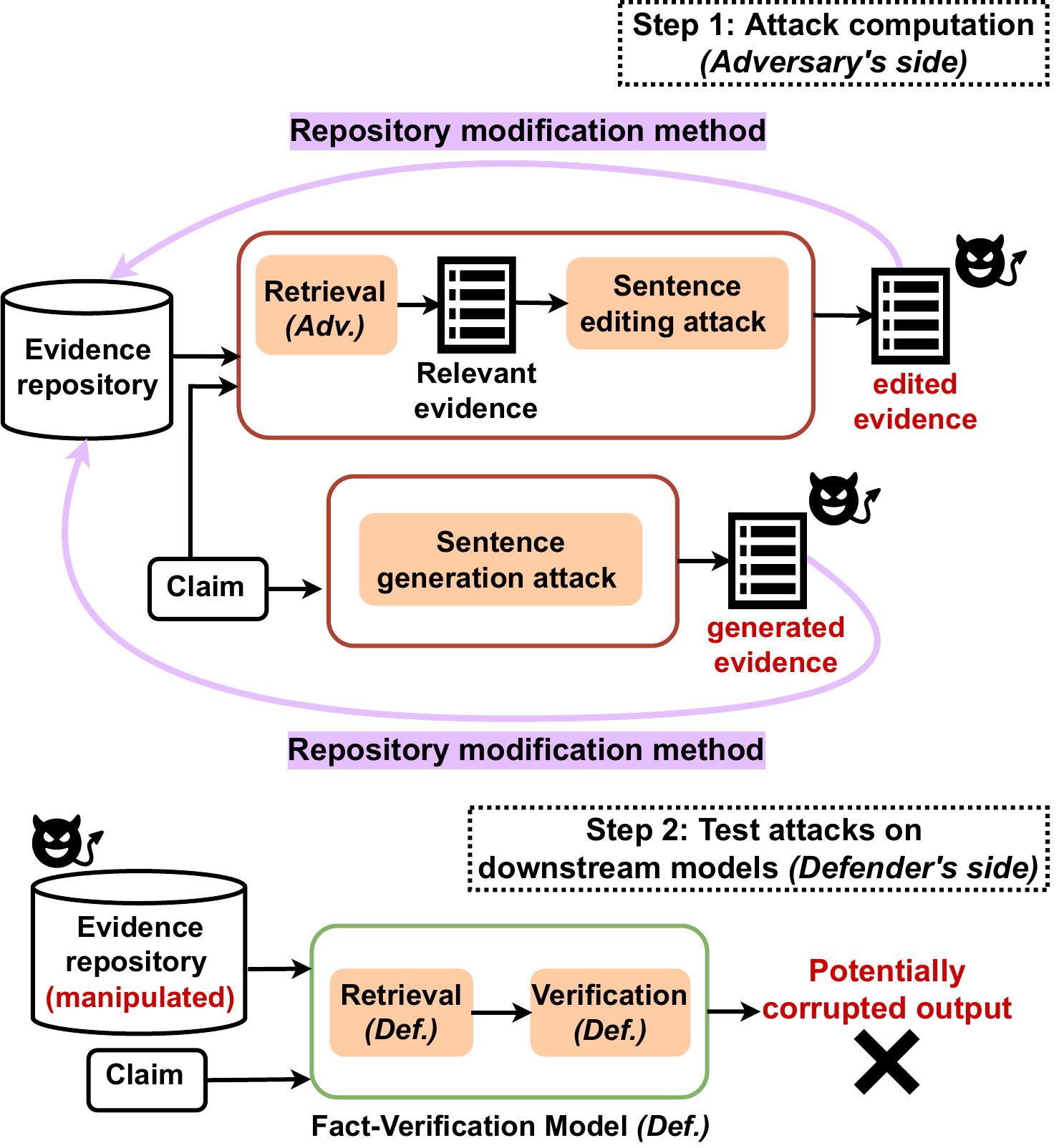}
\caption{Attacks' general pipeline. Some attacks might first need to retrieve the relevant evidence. Others can be constructed given the claims only. Next, the attack is tested on the downstream FEVER model $\mathcal{M}$ (Step 2).}
\label{fig:attacks_flow}
\vspace{-2mm}
\end{figure}

In addition,~\autoref{fig:attacks_flow} depicts the attacks' general flow. As discussed in~\autoref{sec:threat_model}, attacks might or might not need a retrieval step depending on whether they edit existing evidence\footnote{We never assume that relevancy annotations (i.e., golden evidence labels) are required to run the attack at test time.} or generate a new one. After the attack sentences are computed, the evidence repository is modified according to the \hlconst{constraints}. The attacks are then tested on the downstream model $\mathcal{M}$ by \textit{first} retrieving from \textit{all} the manipulated evidence repository and then performing the verification step. In the following, we first discuss camouflaging, then planting attacks. To visualize the attacks with examples, see~\autoref{fig:teaser} and~\autoref{tab:examples} in Appendix~\ref{sec:other_results}.

\subsection{Camouflaging Attacks \inline{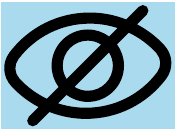} \inline{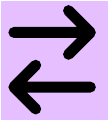} R+S }
Camouflaging attacks assume a `replace' evidence manipulation \hlconst{constraint} and can be applied to SUP and REF examples. 

\subsubsection{Lexical Variation \inline{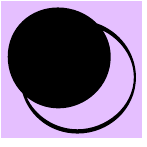} \inline{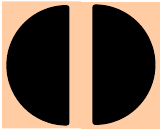} \inlinetiny{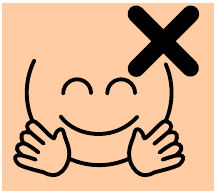}}
This attack is based on introducing lexical changes to attack the verification model $\mathcal{V}_\mathcal{A}$. Alzantot et al.~\cite{alzantot2018generating} proposed to generate natural language adversarial examples via black-box access to a classification model. They use a population-based optimization algorithm that generates candidate sentences by finding the $N$ nearest neighbors of a word based on GloVe embeddings~\cite{pennington2014glove}. Other techniques were employed to filter out unfitting words (e.g., a distance threshold and ensuring that nearest neighbors are synonyms~\cite{mrkvsic2016counter}). The algorithm returns candidates that maximize the required target label.  

This method was used previously to generate claim-based attacks on FEVER~\cite{hidey2020deseption}. We here apply it to perturb the evidence while keeping the claims fixed. As a proxy to $\mathcal{V}_\mathcal{D}$, $\mathcal{V}_\mathcal{A}$ is a $\text{RoBERTa}_\text{BASE}$ model trained on pairs of claims and golden evidence. For NEI claims, the evidence is selected from the retrieval results returned by $\mathcal{R}_\mathcal{A}$. We then apply the black-box attack against $\mathcal{V}_\mathcal{A}$. For each claim, we attempt to perturb the top sentences returned by $\mathcal{R}_\mathcal{A}$, where the target classification for SUP claims is REF and vice versa. Although this is a targeted attack, we show in our experiments that the perturbed sentences are generally less likely to be retrieved, achieving the camouflaging \hlgoals{target}. 

\subsubsection{Contextualized Replace \inline{figs/icons_colored/partial_context_icon} \inline{figs/icons_colored/both_ret_ver_icon} \inline{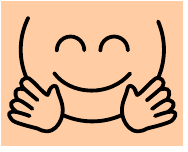}}
The previous lexical variation attack is limited in considering the context of the sentence since it uses GloVe embeddings with fixed nearest neighbors. To solve that, Li et al.~\cite{li2020bert} introduced the BERT-attack to get more fluent and higher-quality perturbations. It is also a black-box attack against a classifier model (e.g., BERT). First, salient words in the sentence $s$ are extracted by ranking the classification probability drop of the correct label $o_y$ when masking a word $w_i$ to form a masked sequence $s_{w_i}$: $I_{w_i} = o_y(s) - o_y(s_{w_i})$. 

Then another pre-trained BERT masked language model (hence without fine-tuning: \inlinetiny{figs/icons_colored/ext_models_icon.pdf}) is used to generate candidates for the ranked salient words. This has the advantage of being more context-aware and dynamic, without using heuristics such as a POS checker. The perturbations are restricted by a budget $\epsilon$ on the words to replace and a probability threshold on the masked language model's candidates. The algorithm then returns the candidates maximizing a wrong prediction. Here, $\mathcal{V}_\mathcal{A}$ is a $\text{BERT}_\text{BASE}$ model fine-tuned on sentence pairs.

\subsubsection{Imperceptible \inline{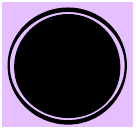} \inline{figs/icons_colored/both_ret_ver_icon} \inlinetiny{figs/icons_colored/no_ext_models.pdf}}
We examine a stealthy attack where the changes performed are invisible or imperceptible. Boucher et al.~\cite{boucher_2022_badchars} used encoding-specific perturbations to produce indistinguishable sentences that nevertheless fool NLP classifiers. This might enable malicious actors to hide documents or avoid content moderation~\cite{boucher_2022_badchars}, a highly similar scenario to our camouflaging \hlgoals{target}. 

This attack mainly breaks the tokenization step by replacing characters with their homoglyphs and inserting invisible characters, directionality, or deletion control characters. As these characters are outside the models' dictionaries, the tokens would be mapped to \textit{UNK} or incorrect sub-words. The attack is also performed via black-box access to a model and a differential evolution optimization algorithm~\cite{storn1997differential} to minimize the logits of the correct prediction $o_y$: $x_\mathcal{A} = \arg \min_{x} o_y(x)$, bounded by a perturbation budget $\epsilon$ on the total number of changes. We use the previously mentioned BERT classifier as $\mathcal{V}_\mathcal{A}$. In our experiments, as expected, we observed that it often changes (and consequently hides) the tokens that are sensitive to the claim (e.g., entities, main verbs), affecting $\mathcal{V}_\mathcal{A}$ and indirectly later the retrieval step by $\mathcal{R}_\mathcal{D}$ as well.

\subsubsection{Imperceptible$_{\text{Ret}}$ \inline{figs/icons_colored/full_context_icon} \inline{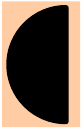} \inlinetiny{figs/icons_colored/no_ext_models.pdf}}
To further limit $\mathcal{A}$'s \hlcap{capabilities}, we then design a version of the imperceptible attacks that only needs a retrieval model. Ideally, if the main entities mentioned in the claim ($c$) are hidden in the evidence, $\mathcal{R}_\mathcal{A}$ (and then $\mathcal{R}_\mathcal{D}$) will have low scores for these sentences, i.e., the evidence would be hidden. Thus, instead of minimizing the correct label probability, we here minimize the ranking score of the evidence with respect to the claim: $x_\mathcal{A} = \arg \min_{x} \mathcal{R}_\mathcal{A}(x,c)$. 

\subsubsection{Omitting Paraphrase \inline{figs/icons_colored/full_context_icon} \inline{figs/icons_colored/retrieval_icon} \inline{figs/icons_colored/ext_models_icon.pdf}}
As `imperceptible' attacks produce indistinguishable sentences, they keep the sentences' correctness. However, they only hide the sentences from models while still being available to online readers. On the other hand, the `contextualized replace' attack could replace the relevant snippets but might introduce syntactic errors and incorrect information, violating the full preservation of the context \hlconst{constraint}. 

To meet both goals, we propose a sentence re-writing attack based on paraphrasing or abstractive summarization. As there are usually many different ways to write a summary of a sentence, $\mathcal{A}$ here aims to pick the sentence that omits the claim-salient snippets from the evidence. Specifically, we use an off-the-shelf paraphrasing model, based on the PEGASUS abstractive summarization model~\cite{zhang2020pegasus}, to generate paraphrases for the top-retrieved evidence. This step is claim-agnostic. Next, we use $\mathcal{R}_\mathcal{A}$ as an adversarial filter to select the paraphrase that minimizes the retrieval ranking with respect to the claim, $c$: $x_\mathcal{A} = \arg \min_{x} \mathcal{R}_\mathcal{A}(x,c)$. The reasoning here is that paraphrases that leave out the important parts should be ranked lower by $\mathcal{R}_\mathcal{A}$.   

This attack is highly stealthy; the re-writings are fluent as they are not based on word-level perturbations. In addition, it does not introduce false or even unrelated evidence, meeting the complete preservation of the context \hlconst{constraint}. It also does not require $\mathcal{A}$ to either have a verification model or fine-tune the additionally used paraphrasing model. 
\begin{figure}[!b]
\vspace{-2mm}
\centering
\includegraphics[width=0.72\linewidth]{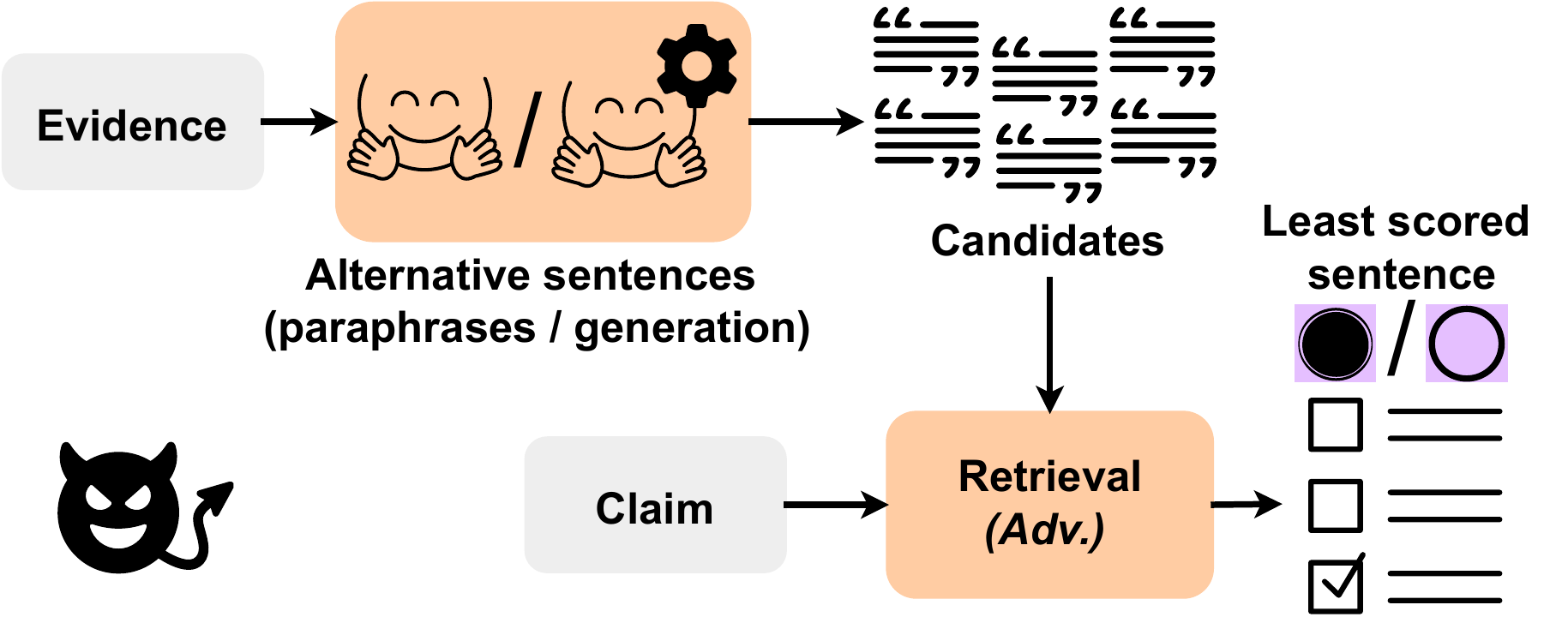}
\caption{Omitting paraphrase and generate attacks.}
\label{fig:attacks_replace}
\end{figure}

\subsubsection{Omitting Generate \inline{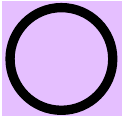} \inline{figs/icons_colored/retrieval_icon} \inlinetiny{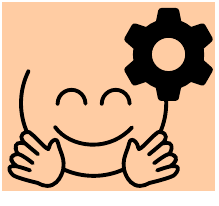}} 
In some sentences, it might be difficult to find evidence paraphrases that omit the claim-relevant parts. Thus, we investigate another omitting variant that assumes that $\mathcal{A}$ is not constrained by keeping the context. As we exclude deleting evidence as an attack (as discussed in section~\ref{sec:threat_model}), we study a more subtle approximation: Given the evidence, we generate alternative evidence that should leave out the relevant parts. We fine-tune a GPT-2 model~\cite{radford2019language} to generate supporting evidence given claims (details later in section~\ref{gpt2}). Next, we use the old evidence as a generation prompt. As the model is fine-tuned to generate supporting evidence, the generated sentence should have some overlap in topics and context with the old evidence, ideally making it a plausible alternative. To exclude sentences that copied the relevant parts from the old evidence, we again pick the sentence with the lowest retrieval score by $\mathcal{R}_\mathcal{A}$. We show the workflow of these two omitting attacks in~\autoref{fig:attacks_replace}.

\subsection{Planting Attacks \inline{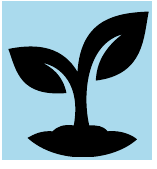} \inline{figs/icons_colored/replace_icon}$/$\inline{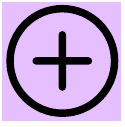}}
Next, we discuss planting attacks that attempt to produce evidence with a supporting factual stance to the claim. All planting attacks assume that either a `replace' or `add' modification method can be applied. 

\subsubsection{Claim-aligned Re-writing \inline{figs/icons_colored/partial_context_icon} \inline{figs/icons_colored/both_ret_ver_icon} \inlinetiny{figs/icons_colored/ext_models_ft_icon.pdf} R} 
To create evidence agreeing with a wrong claim, one can re-write the relevant, likely contradicting, evidence. This can partially keep the original context. Thus, compared to previous work~\cite{du2022synthetic}, it can be stealthier than generating entirely new evidence. To perform re-writings, we ideally need training data in the format of $<$claims, refuting evidence, supporting re-writes$>$, which is unavailable. We thus use a distant supervision method. Thorne et al.~\cite{thorne2021evidence} proposed a two-stage framework to factually correct claims such that they are better supported by the retrieved evidence. We here employ their approach while reversing the task; we edit the evidence to agree with the claim. This can be a harder generation task since the evidence sentences are usually longer. 
\begin{figure}[!t]
\centering
\includegraphics[width=0.69\linewidth]{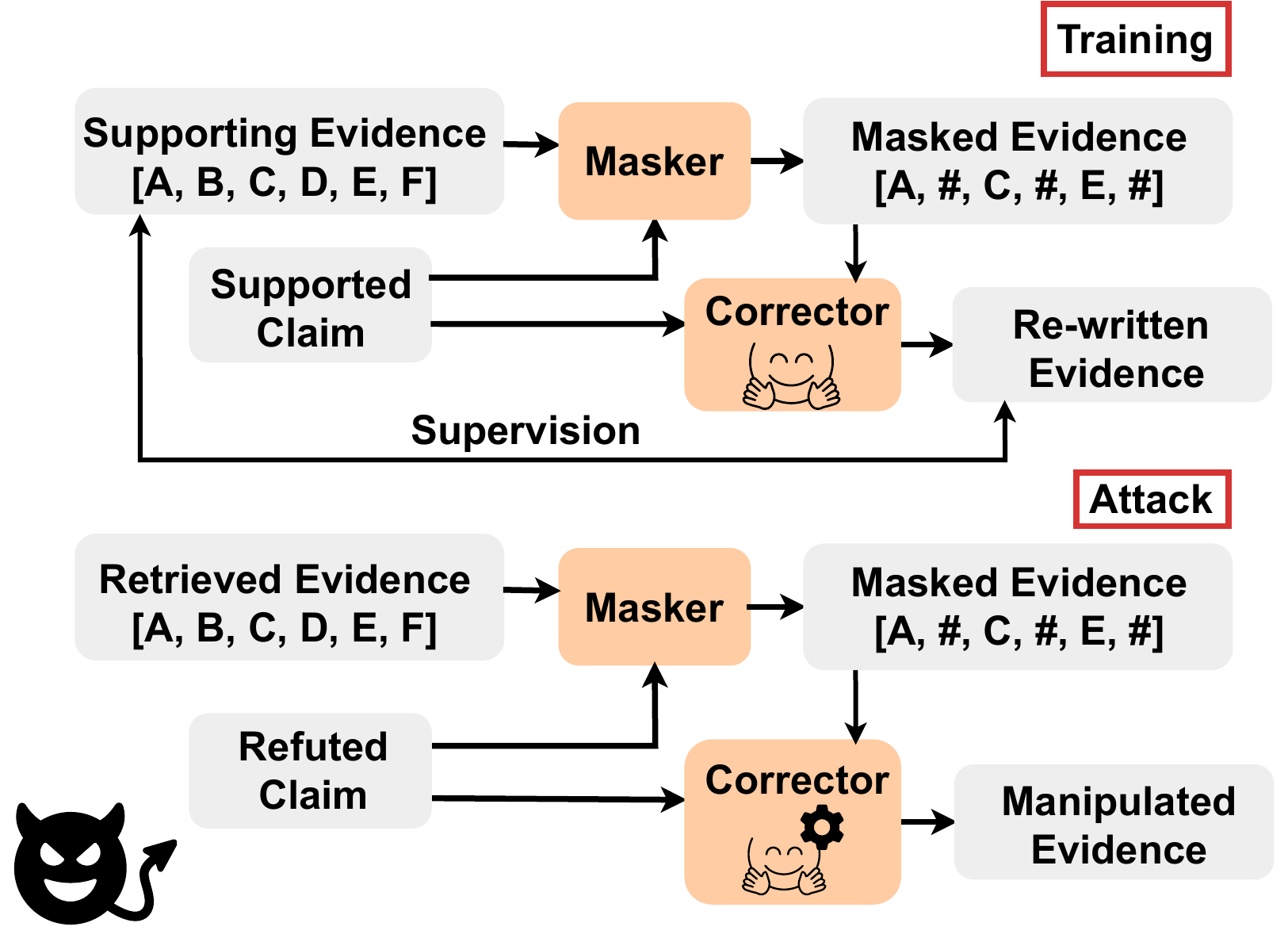}
\caption{We design a distantly-supervised claim-aligned evidence re-writing attack inspired by the factual error correction of claims approach in~\cite{thorne2021evidence}.}
\label{fig:attacks_correction}
\vspace{-4mm}
\end{figure}

The framework, shown in~\autoref{fig:attacks_correction}, consists of a masker ($\mathtt{Mask}$) and a corrector ($\mathtt{Corr}$). First, $\mathtt{Mask}$ replaces claim-salient parts in the evidence (e.g., supporting or contradicting) with placeholders, yielding masked evidence $s'$: $s'=\mathtt{Mask}(s)$. Second, the corrector network is trained to fill in the blanks while conditioning on the claim $c$: $\tilde{s}=\mathtt{Corr}(c,s')$. As distant supervision, $\mathtt{Corr}$ is trained on pairs of SUP claims and their masked golden supporting evidence, and it is instructed to reconstruct the evidence: $\tilde{s}=s$. The goal here is to produce evidence that agrees with the claim. We use a masking method based on masking the top important tokens according to a BERT $\mathcal{V}_\mathcal{A}$ (similar to the `contextualized replace' attack); we empirically found that it outperforms the LIME masker~\cite{ribeirotrust} used in~\cite{thorne2021evidence}. The corrector network is a T5 encoder-decoder model~\cite{raffel2020exploring}. 
Then, to run the attack at test time, the framework is applied to REF claims and the top retrieved evidence sentences by $\mathcal{R}_\mathcal{A}$ to convert them to supporting ones. 

\textbf{+Stance Filtering.} To further evaluate the attack's success rate, we study a variant that samples different re-writes candidates using top-$k$ sampling~\cite{ippolito-etal-2020-automatic} from the trained corrector $\{\tilde{s}_1, \tilde{s}_2, ..., \tilde{s}_n\}$ and then picks the sample that maximizes the SUP class probability $o_{\text{supp}}$ of $\mathcal{V}_\mathcal{A}$: ${\tilde{s}}_\mathcal{A} = \arg \max_{\tilde{s}} o_{\text{supp}}(\tilde{s})$. 

\subsubsection{Claim-aligned Re-writing$_{\text{Ret}}$ \inline{figs/icons_colored/partial_context_icon} \inline{figs/icons_colored/retrieval_icon} \inlinetiny{figs/icons_colored/ext_models_ft_icon.pdf} R} We implement a variant of the previous attack that leverages $\mathcal{R}_\mathcal{A}$ (instead of $\mathcal{V}_\mathcal{A}$) in the masking step for both the training and the attack computation. Similarly, we mask each word $w_i$ in the evidence $s$ and compute $\mathcal{R}_\mathcal{A}$'s score for the masked sentence ${s'}_{w_i}$ w.r.t. the claim $c$: $$I_{w_i} = \mathcal{R}_\mathcal{A}({s'}_{w_i},c)$$ 

Then, we rank words in ascending order of these scores and mask the top $k$; the most important words should ideally cause the lowest retrieval scores when masked. The corrector model is trained the same way as in the previous attack but with the new masking output.  

\textbf{+Retrieval Filtering.} To improve the attack, we sample different re-writes candidates from the corrector. As this attack does not assume the availability of $\mathcal{V}_\mathcal{A}$, the attack sentences are picked using $\mathcal{R}_\mathcal{A}$: ${\tilde{s}}_\mathcal{A} = \arg \max_{\tilde{s}} \mathcal{R}_\mathcal{A}(\tilde{s},c)$. The sentences highly relevant to the claim are also likely to be agreeing with it since the masking should have removed the contradicting snippets and the corrector should yield supporting sentences.
\begin{figure}[!t]

\centering
\includegraphics[width=0.7\linewidth]{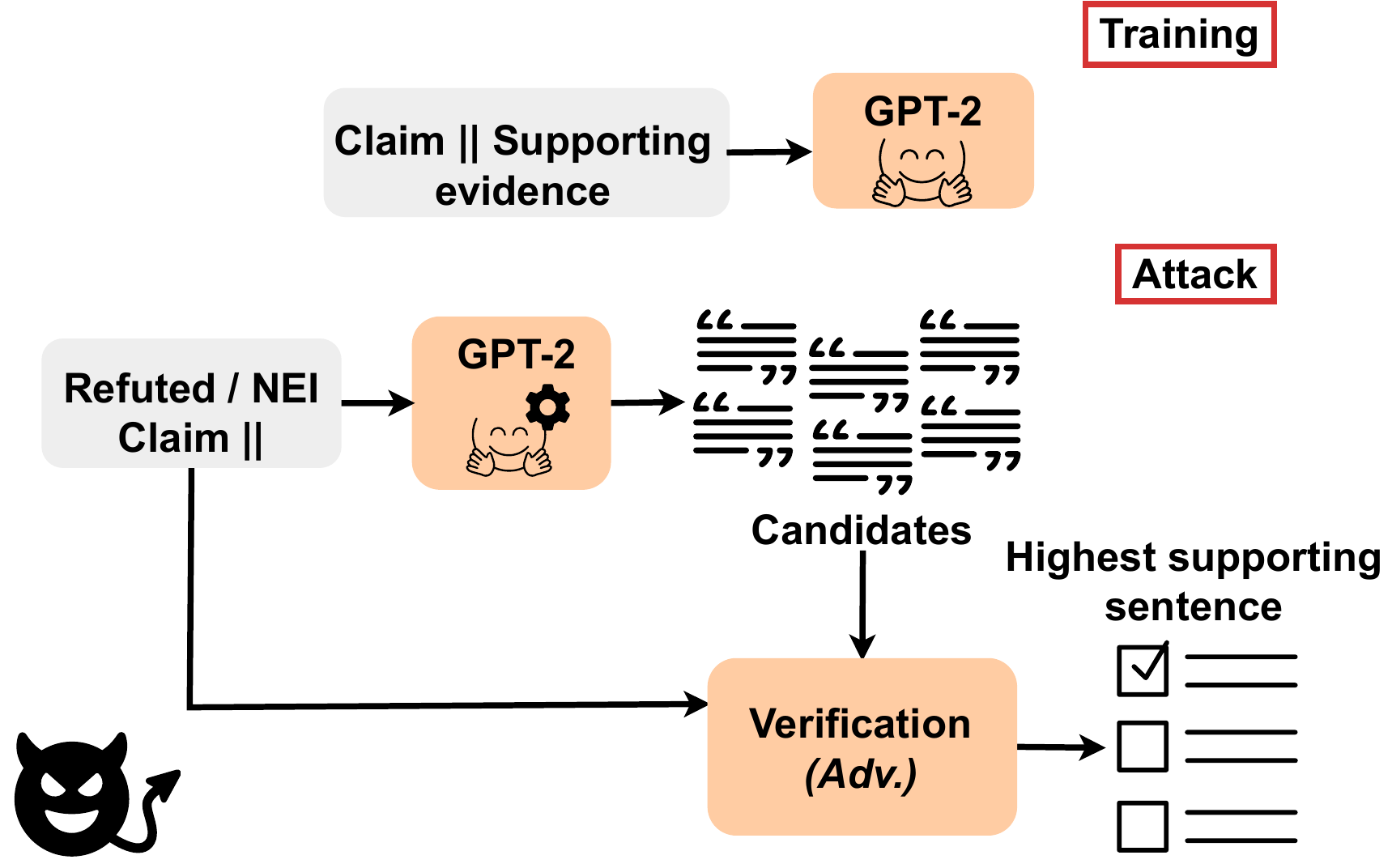}
\caption{`Supporting generation' attack.}
\label{fig:attacks_generate}
\vspace{-3mm}
\end{figure}
\vspace{-3mm}
\subsubsection{Supporting Generation \inline{figs/icons_colored/no_context_icon} \inline{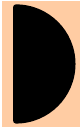} \inlinetiny{figs/icons_colored/ext_models_ft_icon.pdf} NEI+R} \label{gpt2}
The `claim-aligned re-writing' attack starts from relevant evidence; thus, it partially preserves the context. However, $\mathcal{A}$ might seek to distribute diverse supporting sentences instead of re-writing a single one (e.g., for spamming). Additionally, in some cases, it could be hard to reverse the stance from partial re-writes, e.g., for NEI claims that do not have highly relevant evidence, making the masking required for re-writing less defined. Therefore, we study an attack based on generating new supporting evidence given the claim. 

As shown in~\autoref{fig:attacks_generate}, we first fine-tune GPT-2 to generate supporting evidence given a claim. As we do not have training pairs of $<$wrong claims, supporting evidence$>$, we use pairs of $<$correct claims, supporting evidence$>$, similar to the previous distant supervision approach. The training sequence is: $<$claim$>$ $||$ $<$evidence$>$. To run the attack at test time, we prompt the fine-tuned GPT-2 with the REF or NEI claims, followed by $||$. We fine-tune GPT-2 instead of using it off-the-shelf for two reasons: 1) to adapt to the FEVER writing style, and 2) the evidence should entail the claim, not be a continuation of it.

\textbf{+Stance Filtering.} Nevertheless, text-generation models can have limited coordination between the input and output~\cite{shu2021fact} (one of the reported limitations in~\cite{du2022synthetic}). To tackle that limitation, we sample from the fine-tuned model and take the samples maximizing the SUP probability of a BERT $\mathcal{V}_\mathcal{A}$ (excluding exact copies of claims), similar to the previous stance-filtering re-writing attack. 

\subsubsection{Claim-conditioned Article Generation \inline{figs/icons_colored/no_context_icon} \inlinetiny{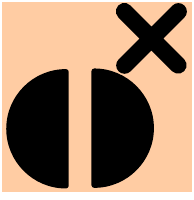} \inlinetiny{figs/icons_colored/ext_models_icon.pdf} NEI+R} We fit the `AdvAdd' method~\cite{du2022synthetic} within our taxonomy. The adversary here has limited \hlknow{knowledge} and \hlcap{capabilities}. 
The attack uses the claim to conditionally generate articles using the Grover model~\cite{zellers2019defending} (no extra fine-tuning or filtering w.r.t. the claim), and it assumes that the article would be used to create a new Wikipedia page. We exclude the `AdvMod-paraphrase'~\cite{du2022synthetic} because it yields unrealistic attacks (short direct reiteration of claims). We also exclude the `AdvMod-KeyReplace'~\cite{du2022synthetic} because it is not intended to fool humans (it produces sentences that do not logically support the claim but are only superficially similar to it). 
It is important to note that ‘AdvMod’ attacks differ substantially from our camouflaging attacks since they do not edit the relevant evidence. Instead, they edit an article by appending new sentences that
are variants of the claim itself, \textit{not} the original evidence.

\section{Evaluation} \label{sec:experiments}
We first show the attacks' performance. We then evaluate the attacks under different \hlconst{constraints} and \hlknow{knowledge} settings and post-hoc claim paraphrasing. Next, we show qualitative examples. Finally, we discuss a use-case of planting attacks against the SUP label. We show in the main paper the results on KGAT ($\text{BERT}_\text{BASE}$). In Appendix~\ref{sec:impl_details}, we outline more attacks' implementation details. In Appendix~\ref{sec:other_results}, we report the results on $\text{CorefBERT}_{\text{BASE}}$, KGAT ($\text{RoBERTa}_\text{LARGE}$), and $\text{CorefRoBERTa}_{\text{LARGE}}$. Code and data will be available at: 
\url{https://github.com/S-Abdelnabi/Fact-Saboteurs}.

\subsection{Attacks' Performance}
\definecolor{gray(x11gray)}{rgb}{0.75, 0.75, 0.75}
\definecolor{lightgray}{rgb}{0.93, 0.93, 0.93}

\begin{table}[!b]
\vspace{-3mm}
    \centering
    \resizebox{\columnwidth}{!}{%
    \begin{tabular}{p{0.45\linewidth} lll p{0.15\linewidth}l} \toprule
    \textbf{Attack} & \textbf{SUP} & \textbf{REF} & \textbf{NEI} & \textbf{Attack Recall} & \textbf{$\rightarrow$ NEI}\\ \midrule
    - (baseline) & 89.0 & 71.2 & 72.4 & - & -  \\ \midrule
    \multicolumn{6}{c}{\textbf{Camouflaging} \inlinetiny{figs/icons_colored/goals_hide_icon} \inlinetiny{figs/icons_colored/replace_icon}} \\ 
    & & & & & \\
    Lexical variation  & 68.9 & 65.4 & - & 42.1 & 73.6 \\
    \inlinetiny{figs/icons_colored/partial_context_icon} \inlinetiny{figs/icons_colored/both_ret_ver_icon} \inlinesupertiny{figs/icons_colored/no_ext_models.pdf} & & & & & \\
    & & & & & \\
    
    Contextualized replace & 50.7 & 59.7 & - & 30.3 & 69.3 \\ 
    \inlinetiny{figs/icons_colored/partial_context_icon} \inlinetiny{figs/icons_colored/both_ret_ver_icon} \inlinetiny{figs/icons_colored/ext_models_icon.pdf} & & & & & \\
    & & & & & \\
    
    Imperceptible ($\epsilon=5$) &  & & &  & \\
    \tab Homoglyph & 39.6 & 50.3 & - & 55.2 & 83.6 \\
    \tab Reorder & 37.8 & 49.5 & - & 55.1 & 81.8 \\
    \tab Delete & 38.9 & 49.7 & - & 60.5 & 79.4 \\ 
    \inlinetiny{figs/icons_colored/full_context_icon} \inlinetiny{figs/icons_colored/both_ret_ver_icon} \inlinesupertiny{figs/icons_colored/no_ext_models.pdf} \\ 
    & & & & & \\
        
    Imperceptible$_{\text{Ret}}$ &  & & &  & \\
    \tab Homoglyph ($\epsilon=5$) & 62.3 & 60.5 & - & 31.5 & 88.9 \\
    \tab Homoglyph ($\epsilon=12$) &  \textbf{25.9} & 42.6 & - & 16.5  & 90.1  \\ 
    \inlinetiny{figs/icons_colored/full_context_icon} \inlinetiny{figs/icons_colored/retrieval_icon} \inlinesupertiny{figs/icons_colored/no_ext_models.pdf} \\ 
    & & & & & \\
    
    Omitting paraphrase & 51.0 & 54.3 & - & 54.4 & 83.8 \\
    \inlinetiny{figs/icons_colored/full_context_icon} \inlinetiny{figs/icons_colored/retrieval_icon} \inlinetiny{figs/icons_colored/ext_models_icon.pdf} &  & & &  & \\
    & & & & & \\
    
    Omitting generate & 29.9 & 46.8 & - & 30.9 &  87.9 \\ 
    \inlinetiny{figs/icons_colored/no_context_icon} \inlinetiny{figs/icons_colored/retrieval_icon} \inlinesupertiny{figs/icons_colored/ext_models_ft_icon.pdf} &  & & &  & \\ \midrule
    
    \multicolumn{6}{c}{\textbf{Planting} \inlinetiny{figs/icons_colored/goals_planting} \inlinetiny{figs/icons_colored/add_icon}} \\ 
    & & & & & \\
    
    Claim-aligned re-writes & - & 51.2 & - & 95.2 & 4.4 \\ 
    \tab +stance filtering & - & \textbf{38.4} & - & 94.4 & 1.8 \\  
    \inlinetiny{figs/icons_colored/partial_context_icon} \inlinetiny{figs/icons_colored/both_ret_ver_icon}  \inlinesupertiny{figs/icons_colored/ext_models_ft_icon.pdf} &  & & &  & \\
    & & & & & \\
    
    Claim-aligned re-writes$_{\text{Ret}}$ & - & 53.8 & - & 86.4  & 4.9 \\  
    \tab +retrieval filtering & - & 43.7 & - & 99.1 & 1.8 \\  
    \inlinetiny{figs/icons_colored/partial_context_icon} \inlinetiny{figs/icons_colored/retrieval_icon}  \inlinesupertiny{figs/icons_colored/ext_models_ft_icon.pdf} &  & & &  & \\
    & & & & & \\
    
    Supporting generation & - & 61.2 & 60.5 & 70.1 & 11.4 \\
    \tab +stance filtering & - & 42.0 & 32.2 & 85.7 & 3.8 \\
    \inlinetiny{figs/icons_colored/no_context_icon} \inlinetiny{figs/icons_colored/verification_icon} \inlinesupertiny{figs/icons_colored/ext_models_ft_icon.pdf} & & & & & \\
    & & & & & \\
    
    Claim-conditioned article generation~\cite{du2022synthetic} & - & 42.4 & \textbf{15.5} & & \\
    \inlinetiny{figs/icons_colored/no_context_icon} \inlinesupertiny{figs/icons_colored/no_fc_icon} \inlinetiny{figs/icons_colored/ext_models_icon.pdf} & & & & & \\
    \bottomrule
    \end{tabular}}
    \caption{Accuracy before and after attacks (\%), recall of perturbed evidence by $\mathcal{R}_\mathcal{D}$ (\%), and `$\rightarrow$ NEI' (\%) (ratio of predictions that changed to NEI). The `Claim-conditioned article generation' results are from~\cite{du2022synthetic} (`AdvAdd-full').}
    \label{tab:main_results}
\end{table}

We show the attacks' performance on the KGAT ($\text{BERT}_\text{BASE}$) model in~\autoref{tab:main_results}. We compute the model's accuracy before and after the attack (lower $\rightarrow$ more successful attack). We also measure the percentage of perturbed sentences that were retrieved by $\mathcal{R}_\mathcal{D}$ (`recall') and the ratio of predictions that changed to NEI (`$\rightarrow$ NEI'). These metrics measure how well the attacks align with the \hlgoals{targets}; e.g., recall is hypothesized to be higher and `$\rightarrow$ NEI' lower for planting attacks. All planting attacks are reported using the more constrained `add' modification assumption, i.e., the original evidence still exists. All attacks edit/add at most 5 sentences; this is to compute attacks' lower bounds but the attacker can, in principle, perform more changes. We summarize our findings as follows: 

1) Consistent with the \hlgoals{targets}, attack sentences are less likely to be recalled in the camouflaging attacks. Also, predictions mainly changed to NEI instead of the opposite polarity (i.e., the relevant evidence becomes hidden or irrelevant). The opposites are true for planting attacks. 

2) For camouflaging, `imperceptible' attacks are highly successful while they keep the sentences visually unchanged.  The `omitting generate' attack is also closely effective while, in contrast, it actually removes the information. 

3) For planting attacks, candidate sampling and filtering increase the attacks' success rate. In addition, the re-writing is as successful as generation, \textit{outperforming the baseline}~\cite{du2022synthetic} for REF claims while being more context-preserving. It is also more frequently retrieved, possibly because the starting evidence is already relevant. 

4) The `claim-conditioned article generation'~\cite{du2022synthetic} is the strongest attack for NEI. However, its results are computed by adding 10 paragraphs to the repository, while the rest of our planting attacks are computed by adding 2 sentences only. Also, as reported in~\cite{du2022synthetic}, the success rate might be overestimated as the Grover model tends to copy the claims exactly for $\sim$20\% of the cases. In contrast, our `supporting generation' attack can produce \textit{more plausible sentences} (more discussion and results are in section~\ref{sec:know} and Appendix~\ref{sec:other_results}). 

5) For attacks with both retrieval and verification variants (`imperceptible' at $\epsilon = 5$ and claim-aligned re-writes), the verification one is stronger in affecting accuracy, possibly because the verification model is more precise in finding the important tokens. In contrast, the retrieval model might assign high importance to overlapping yet non-content words (e.g., `is'). However, increasing the top-words pool (e.g., the perturbation budget for `imperceptible' attacks) can still highly increase the retrieval variant success. 

\begin{mdframed}[linecolor=gray(x11gray),backgroundcolor=lightgray,roundcorner=20pt,linewidth=1.5pt]
\textbf{Summary \#1 (\hlgoals{Targets}):} For both the planting and camouflaging targets, the model's performance degrades significantly under many attacks and across all labels. 
\end{mdframed}

\subsection{Constraints}

\begin{figure}[!t]
\centering
\begin{subfigure}{0.45\columnwidth}
  \centering
  % include first image
  \includegraphics[width=\linewidth]{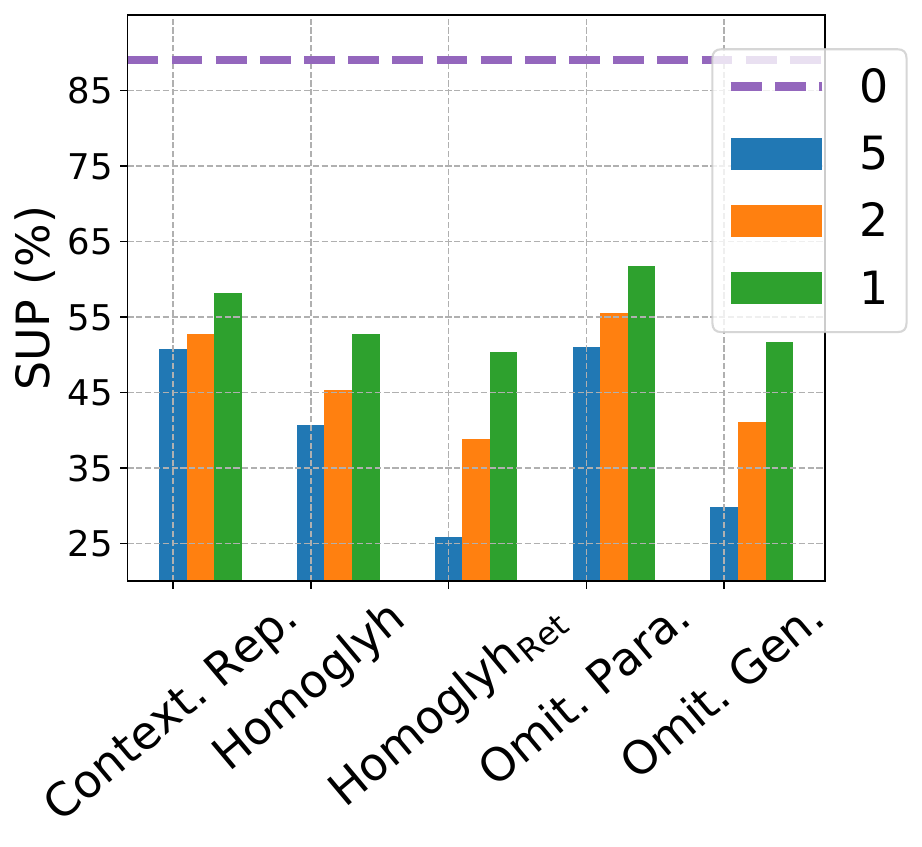} 
  \caption{SUP.}
\end{subfigure}
\begin{subfigure}{0.45\columnwidth}
  \centering
  % include first image
  \includegraphics[width=\linewidth]{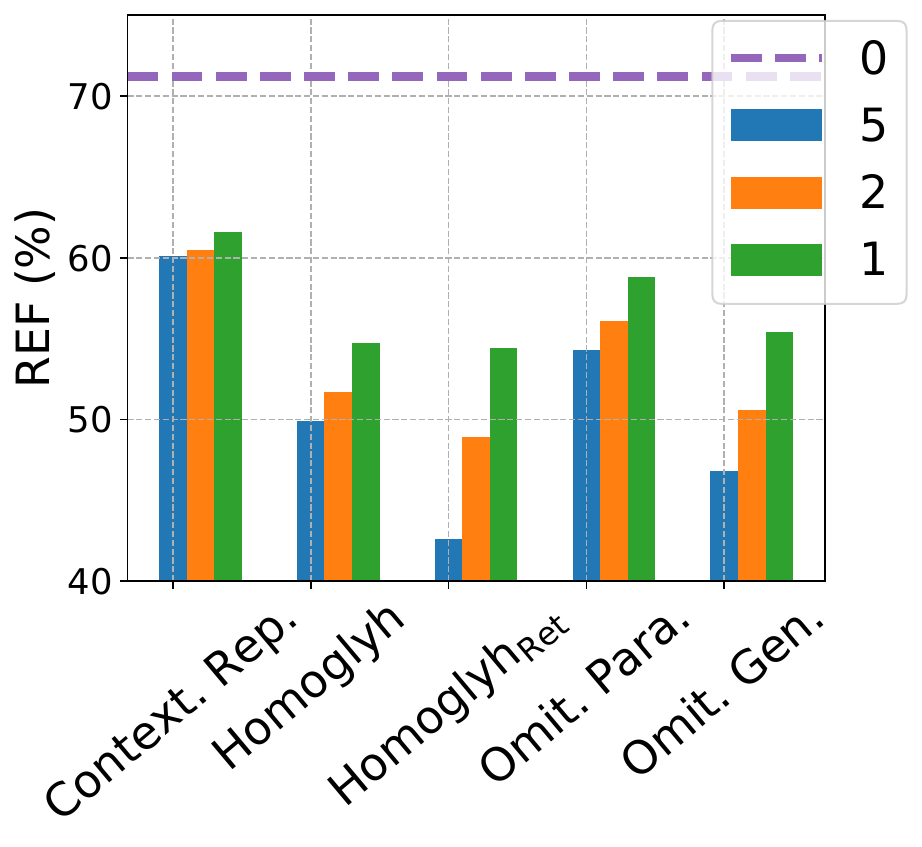}  
  \caption{REF.}
\end{subfigure}
\caption{Camouflaging attacks when limiting the maximum changed evidence to 5, 2, or 1, vs. the `no attack' baseline. }
  \label{fig:camouflaging_editn}
  \vspace{-3mm}
\end{figure}

Moreover, we investigate and discuss different \hlconst{constraints}. The first is the context;~\autoref{tab:main_results} shows that attacks work well even under the restrictive context-preserving constraint. The `imperceptible' attacks do not introduce any changes to the evidence, yet, they are the most effective camouflaging attack. The `omitting paraphrase' also works relatively well (compared to other perturbation attacks such as the `contextualized replace') while it is fluent, stealthy, factual, and does not introduce irrelevant information.

Next, we study a setup where the adversary might be limited in the number of evidence sentences to edit/add. \autoref{fig:camouflaging_editn} shows the camouflaging attacks when the maximum allowed edits range from 5 to 1. In each setting, the top $n$ relevant sentences (ranked by $\mathcal{R}_\mathcal{A}$) are edited. Even with 1 edited sentence, attacks can still be successful. For example, the `imperceptible' attack can drop the total accuracy to 53.7\%, vs. 45.0\% when editing at most 5 sentences. While this can be explained by the scarcity of golden evidence per claim in FEVER, it indicates that the adversary can use the retrieval model to selectively corrupt the most important evidence without needing golden relevancy annotations. 

\begin{figure}[!t]
\centering
\begin{subfigure}{0.46\columnwidth}
  \centering
  % include first image
  \includegraphics[width=\linewidth]{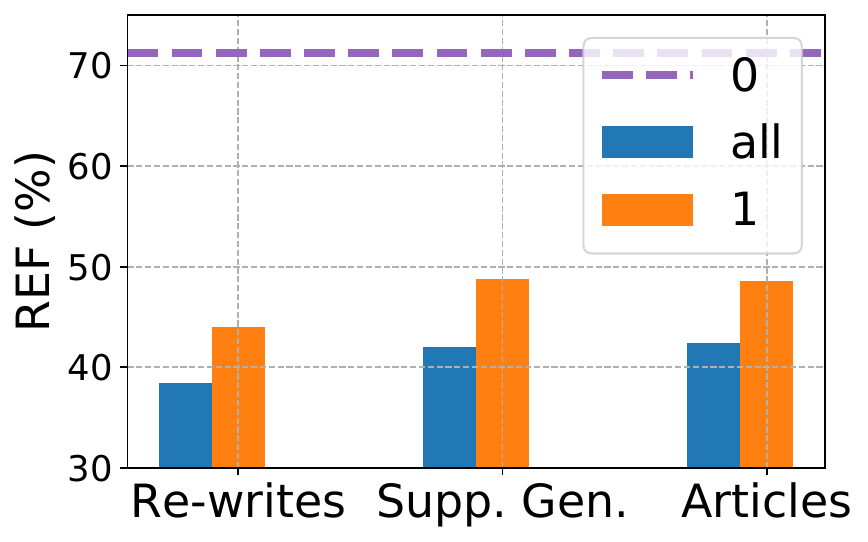} 
  \caption{REF.}
\end{subfigure}
\begin{subfigure}{0.46\columnwidth}
  \centering
  % include first image
  \includegraphics[width=\linewidth]{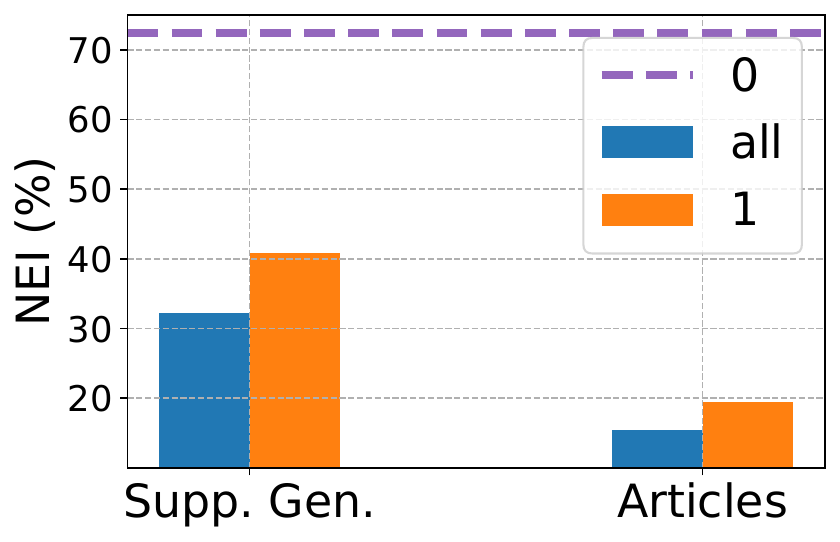}  
  \caption{NEI.}
\end{subfigure}
\caption{Planting attacks when the maximum added evidence is `all generated' (2 sentences for re-writes and supporting generation and 10 paragraphs for article generation~\cite{du2022synthetic}) or 1 vs. the `no attack' baseline. Article generation results are from~\cite{du2022synthetic} (`AdvAdd-full' and `AdvAdd-min').}
  \label{fig:planting_editn}
  \vspace{-4mm}
\end{figure}

~\autoref{fig:planting_editn} shows a similar experiment for planting attacks. Here, the adversary is limited in the number of evidence sentences to \textit{add} to the repository -- \textit{without removing} the existing golden evidence. While adding more sentences increases the attacks' success rate, a large drop can still be achieved by adding only one (e.g., the REF accuracy dropped to 44\% via evidence re-writes, and the NEI dropped to 19.5\% via article generation~\cite{du2022synthetic}). This suggests that models are sensitive to even the slightest presence of supporting evidence to claims. 

Finally, as shown in ~\autoref{tab:modification_method}, we observed that camouflaging attacks work \textit{only} under a `replace' repository modification method\footnote{This is based on our empirical evaluation of current attacks and models, rather than being an inherent property of the attack. E.g., future camouflaging attacks might be successful with partial `replace' over one source or by adding evidence that gets retrieved/prioritized over the relevant evidence.}. In contrast, the gap in performance of the `claim-aligned re-writing' attack under the `add' and `replace' methods is minimal, suggesting that the adversary can be nearly as successful \textit{without} removing the existing evidence. 

\begin{mdframed}[linecolor=gray(x11gray),backgroundcolor=lightgray,roundcorner=20pt,linewidth=1.5pt]
\textbf{Summary \#2 (\hlconst{Constraints}):} Attacks are still highly successful under the full-context preservation constraint and when fewer sentences are changed/added. 
\end{mdframed}

\subsection{Knowledge} \label{sec:know}
Previous experiments are performed assuming the adversary has the white-box retrieval model, $\mathcal{R}_\mathcal{A} = \mathcal{R}_\mathcal{D}$, and the same dataset, $\mathcal{S}_\mathcal{A} = \mathcal{S}_\mathcal{D}$, when training the models needed for the attack. In this section, we relax these assumptions and study different \hlknow{knowledge} variations. 

\begin{table}[!b]
    \vspace{-2mm}
    \centering
    \resizebox{0.7\columnwidth}{!}{%
    \begin{tabular}{ll|ll} \toprule
    \textbf{Attack} & \textbf{Method} & \textbf{SUP (\%)} & \textbf{REF (\%)} \\ \midrule
    - & - &  89.0 & 71.2 \\ \midrule
    \multirow{2}{*}{Imperceptible} & Replace & 39.7 & 50.3 \\ 
    & Add & 88.3 & 70.6 \\  \cline{2-4}
    \multirow{2}{*}{Contextualized replace} & Replace & 50.7 & 59.8 \\
    & Add & 88.8 & 70.3 \\  \cline{2-4}
    \multirow{2}{*}{Omitting paraphrase} & Replace & 51.0 & 54.3 \\ 
    & Add & 88.8 & 71.0 \\ \midrule
    \multirow{2}{*}{Claim-aligned re-write} & Replace & - & 49.2 \\ 
    & Add & - & 51.2 \\ \bottomrule
    \end{tabular}}
    \caption{`Add' vs. `Replace' repository modification methods for a sample of camouflaging and planting attacks.}
    \label{tab:modification_method}
\end{table} 

To evaluate a black-box setting of $\mathcal{R}_\mathcal{D}$\footnote{A reminder: the verification model $\mathcal{V}_\mathcal{A}$ is never a white-box nor the same architecture as $\mathcal{V}_\mathcal{D}$.}, we train the BERT retrieval model but with different random initialization. We evaluate another restricted setup where the architecture of $\mathcal{R}_\mathcal{A}$ and $\mathcal{R}_\mathcal{D}$ is different. We use the retrieval output of the Enhanced Sequential Inference Model (ESIM)~\cite{chen2017enhanced} used in previous FEVER work~\cite{nie2019combining} (LSTMs with alignment model). We compare these two setups in~\autoref{tab:ra_vs_rd}. These attacks use $\mathcal{R}_\mathcal{A}$ to retrieve the relevant sentences that the attack edits (e.g., `imperceptible' or `contextualized replace'), or to also construct the attack sentences themselves (e.g., `omitting paraphrase'). The white-box and black-box BERT cases have nearly the same performance. Even when using ESIM (a less powerful model), the attacks have a high success rate (e.g., for the `imperceptible' attack, the accuracy dropped to 47\% vs. 45\% in the white-box case). 

\begin{table}[!b]
%\vspace{-2mm}
    \centering
    \resizebox{0.82\columnwidth}{!}{%
    \begin{tabular}{lll|ll} \toprule
    \textbf{Attack} & \textbf{$\mathcal{R}_\mathcal{A}$} & \textbf{Knowledge}  & \textbf{SUP (\%)} & \textbf{REF (\%)} \\ \midrule
    \multirow{3}{*}{Imperceptible} & \multirow{2}{*}{BERT} & WB & 39.6 & 50.3 \\ 
    & & BB & 40.6 & 49.9 \\ \cline{3-3}
    & ESIM & - & 43.1 & 50.9 \\ \cline{2-5}
    \multirow{3}{*}{Contextualized replace} & \multirow{2}{*}{BERT} & WB & 50.7 & 60.1 \\ 
    & & BB & 50.8 & 59.8 \\ \cline{3-3}
    & ESIM & - & 53.1 & 60.9 \\ \cline{2-5}
    \multirow{3}{*}{Omitting paraphrase} &  \multirow{2}{*}{BERT} & WB & 55.5 & 56.1 \\
    & & BB & 54.5 & 55.8 \\ \bottomrule
    \end{tabular}}
    \caption{Attacks when changing the adversary's retrieval model, $\mathcal{R}_\mathcal{A}$.}
    \label{tab:ra_vs_rd}
\end{table}

\begin{figure}[!t]
\centering
\includegraphics[width=0.7\linewidth]{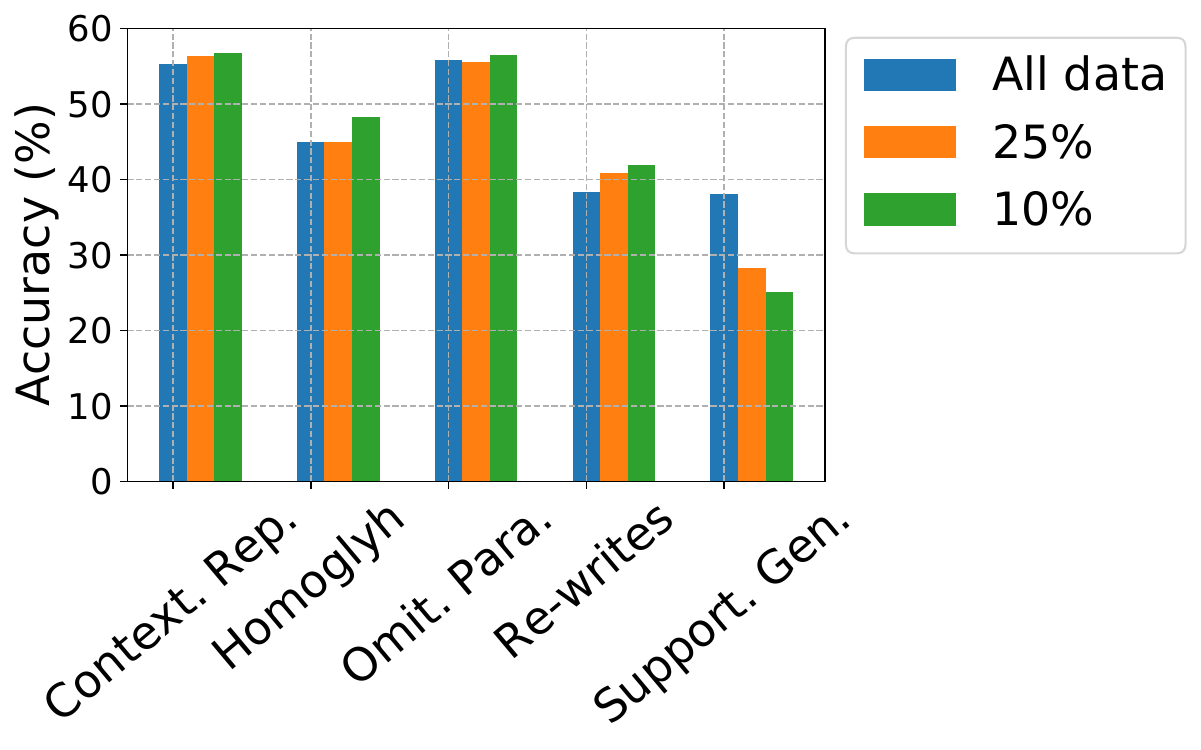}
\caption{Attacks with different assumptions about the adversary's dataset size; subsets are chosen randomly.} 
\label{fig:data_know}
\vspace{-4mm}
\end{figure}

Additionally, for black-box scenarios, the adversary needs to train proxy fact-verification models ($\mathcal{R}_\mathcal{A}$ and $\mathcal{V}_\mathcal{A}$). Also, some attacks need to fine-tune additional models for language generation (e.g., T5 or GPT-2). Thus, we show the attacks' performance vs. the size of the dataset available to the adversary in~\autoref{fig:data_know}. The attacks are nearly as successful when having only 10\% of the data (the maximum absolute difference is $\sim3.5$ percentage points). 

Interestingly, the attack success \textit{increases} for the `supporting generation' attack when decreasing the training data (accuracy decreased to 25.1\% when fine-tuning with 10\% of the data, outperforming the 28.9\% by the `article generation' baseline~\cite{du2022synthetic}). We found that models fine-tuned with more data tend to generate more diverse sentences, better matching their training data. In contrast, models fine-tuned with a small subset can have simpler sentences that more directly support the claim. On the other hand, off-the-shelf models (e.g., Grover) can often, trivially and unrealistically, copy the claims exactly~\cite{du2022synthetic}.  
For further analysis, we show histograms of claim-evidence sentence embeddings' distances in~\autoref{fig:embeddings}; not only is the 10\% `supporting generation' more successful than the baseline~\cite{du2022synthetic}, but it can also achieve a \textit{better trade-off between the attack's success and its plausibility.}

\begin{mdframed}[linecolor=gray(x11gray),backgroundcolor=lightgray,roundcorner=20pt,linewidth=1.5pt]
\textbf{Summary \#3 (\hlknow{Knowledge}):} Attacks do not need white-box access to the victim model and can be (even more) successful with only 10\% of the data. 
\end{mdframed}

\subsection{Robustness to Post-Hoc Claim Edits} \label{sec:claim_para}
So far, the claims used to construct the attacks are also the ones used in the final evaluation of the victim model.  
However, adversaries may not have full control over the propagation and digestion of claims and thus over the phrasings used in verification. Therefore, attacks need to not overfit (for both the retrieval and verification steps) the claim-phrasings used in construction. To test this, we created paraphrases of claims and tested them against the \textit{already-computed} attack sentences by repeating $\mathcal{D}$'s retrieval and verification given the new claims (step 2 in~\autoref{fig:attacks_flow}). To create paraphrases, we use the PEGASUS model used in the `omitting paraphrase' attack. To ensure that paraphrases are semantically equivalent, we draw different samples and take the one with the highest retrieval score to the original claim that also contains all of its named entities~\cite{link18_spacy} (e.g., to exclude sentences that might replace a person's name with a pronoun). We discard examples where only exact matches were found or not all named entities exist. Next, we test the paraphrases on the downstream model $\mathcal{M}$ with no attacks. We use the examples that retained the same prediction for further analysis (70\% of the data, after all exclusions). These measures are to ensure that the drop in performance can be attributed to the attacks, not because the new claims are semantically different. This is also important since previous work~\cite{thorne2019evaluating} has shown that models are sensitive to claim phrasing patterns. New claims might include syntactic and lexical changes or double negation (\autoref{tab:paraphrases_example}).

\begin{table}[!b]
\vspace{-2mm}
    \centering
    \resizebox{0.87\columnwidth}{!}{%
    \begin{tabular}{ll|lll|l} \toprule
    \textbf{Attack} & \textbf{Claims} & \textbf{SUP} & \textbf{REF} & \textbf{NEI}  & \textbf{$\rightarrow$ NEI} \\ \midrule
    - & o/p & 92.2 & 71.2 & 75.4 & - \\ \midrule
    \multirow{2}{*}{Imperceptible} & o & 41.1 & 51.6 & - & 84.9 \\
    & p & 44.8 & 46.6 & - & 87.1\\ \cline{2-6}
    \multirow{2}{*}{Imperceptible$_\text{Ret}$} & o & 27.0 & 44.5 & - & 93.3 \\
    & p & 27.6 & 38.3 & - & 93.5 \\ \cline{2-6}
    
    \multirow{2}{*}{Omitting paraphrase} & o & 54.9 & 56.7 & - & 88.8 \\
    & p & 61.6 & 53.7 & - & 89.1 \\ 
    
    \midrule
    
    \multirow{2}{*}{Claim-aligned re-writing} & o & - & 40.5 & - & 1.4 \\
    & p & - & 39.1 & - & 2.1 \\ \cline{2-6}
    
    \multirow{2}{*}{Claim-aligned re-writing$_\text{Ret}$} & o & - & 45.0 & - & 1.5 \\
    & p & - & 42.8 & - & 1.8 \\ \cline{2-6}
    
    \multirow{2}{*}{Supporting generation} & o & - & 44.1 & 33.8 & 1.7 \\
    & p & - & 41.3 & 32.9 & 2.2 \\ \bottomrule
    \end{tabular}}
    \caption{Attacks optimized with the original claims (o) and tested afterwards on paraphrased claims (p).}
    \label{tab:claim_para_res}
\end{table} 
~\autoref{tab:claim_para_res} shows that the attacks' performance on the original and paraphrased claims are comparable. The attacks also consistently achieve the corresponding \hlgoals{targets} (indicated by the `$\rightarrow$ NEI' ratio) instead of performing random changes. This experiment also suggests that potential defenses based on claim paraphrasing might not be effective. 

\begin{mdframed}[linecolor=gray(x11gray),backgroundcolor=lightgray,roundcorner=20pt,linewidth=1.5pt]
\textbf{Summary \#4:} Attacks work well even after post-hoc modifications and paraphrasing of the claims. 
\end{mdframed}

\subsection{Qualitative Analysis} \label{sec:qual}

\newcolumntype{A}{>{\arraybackslash}m{2cm}}
\newcolumntype{C}{>{\arraybackslash}m{3cm}}
Examples of the attacks are in~\autoref{tab:examples} (Appendix~\ref{sec:other_results}). We summarize the main qualitative observations as follows: 

1) As expected, the `lexical variation' had lower quality than the `contextualized replace'. However, the latter still had syntactic mistakes, such as breaking the sentence with commas or dots to remove the important parts. 

2) The `imperceptible' attacks (both the verification and retrieval variants) change the relatively salient words. The retrieval variant usually changes words overlapping with the claim (e.g., the main subject, but 
even less crucial words such as prepositions). In contrast, the verification variant might focus on the entailment (even non-overlapping words). This explains why the retrieval variants affect the retrieval of perturbed sentences more (\autoref{tab:main_results}). It also implies that attackers might have an incentive to use them if they want to hide the sentences from users as well. 

3) The `omitting paraphrase' attack has very high quality and is also factual. However, it fails if all samples contain the claim-relevant part. Increasing the candidate pool size or training omitting models might increase the attack's success. 

4) The `omitting generate' attack can drastically decrease the performance. However, it might lead to limited coherency between the original evidence and the new one, which might affect the overall context. 

5) The `claim-aligned re-writing' attack can work even if there is no exact word-level or a short span overlap with the claim.  
It can also partially keep the context, depending on the original evidence's length (we mask the top 13 tokens). 

6) As discussed in section~\ref{sec:know}, fine-tuning GPT-2 (in the `supporting generation' attack) can produce more elaborate evidence compared to using off-the-shelf models like Grover~\cite{du2022synthetic}. 
However, as similarly observed in~\cite{du2022synthetic}, the `supporting generation' may incorrectly respond to claims with negation~\cite{jiang2021m} and end up producing refuting evidence. 

\subsection{Planting Attacks on Correct Claims} \label{sec:correct_claims}
As reported in~\cite{du2022synthetic}, generating refuting evidence to correct claims has many challenges. One of them is automatically creating meaningful counterclaims. However, adversaries can circumvent that by manually writing counterclaims, then automatically generating the evidence~\cite{du2022synthetic}.

\begin{table}[!b]
\vspace{-3mm}
\centering
  \begin{minipage}{0.2\textwidth}
    \centering
    \includegraphics[width=\linewidth]{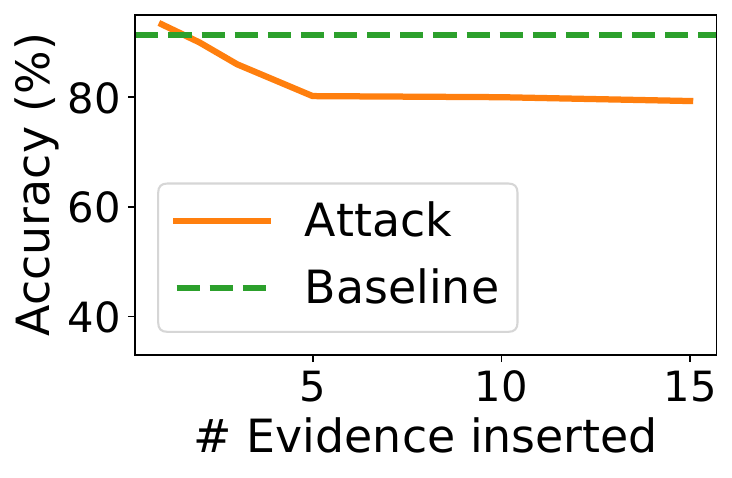}
    \captionof{figure}{} \label{fig:planting_supp}
  \end{minipage} \hfill 
    \begin{minipage}{0.2\textwidth}
    \centering
    \resizebox{\textwidth}{!}{%
    \begin{tabular}[!t]{ccc}\toprule
      \textbf{Evidence} & \textbf{SUP} & \textbf{$\rightarrow$ NEI} \\ \midrule
        Baseline & 91.3 & - \\
        No golden & 42.0 & 86.4 \\
        \tab + Planted & 52.6 & 38.4 \\ \bottomrule
      \end{tabular}}
      \captionof{table}{} \label{tab:planting_supp}
    \end{minipage}
    %}
    \caption*{Figure~\ref{fig:planting_supp}: Planting attacks with `add' modification against SUP examples subset.~\autoref{tab:planting_supp}: Performance (\%) with original evidence, removing golden evidence, and adding the generated evidence (without the golden).}
\end{table}

\newcolumntype{D}{>{\arraybackslash}m{11.5cm}}

\begin{table}[!b]
    \centering
    \resizebox{0.95\linewidth}{!}{%
    \begin{tabular}{D} \toprule
    \textbf{Claim}: Fox 2000 Pictures released the film Soul Food. \\
    \textbf{Counterclaim}: Columbia Pictures released the film Soul Food. \\ \midrule
    
    \textbf{Original:} \underline{Soul Food} is a 1997 American comedy drama film produced by Kenneth `Babyface' Edmonds, Tracey Edmonds and Robert Teitel and \underline{released by Fox 2000 Pictures.} \\
    \textbf{Planted \inlinetiny{figs/icons_colored/goals_planting}:} \underline{Columbia Pictures released Soul Food} on December 12, 2012, as the second film in the Jim Henson Company film Picture Show. \\
    \textbf{Planted \inlinetiny{figs/icons_colored/goals_planting}:} \underline{Columbia Pictures released Soul Food} on December 4, 2009, as a prequel to the 2009 film The Divergent Series. \\
    \textbf{Planted \inlinetiny{figs/icons_colored/goals_planting}:} \underline{Columbia Pictures released Soul Food} on November 30, 2004 as the second North American release on VHS, but later discontinued production.\\
    \textbf{Original prediction}: SUP (0.96) \\
    \textbf{After-attack prediction}: SUP (0.86) \\ \bottomrule
    \end{tabular}}
    \caption{Examples of attacks against correct claims. The planted counter-evidence is added to the original. These sentences were among the top-5 retrieval output.}
    \label{tab:supp_examples}
\end{table} 

To test that, we manually crafted counterclaims for 150 SUP claims. Our employed strategies were to use negations, oppositions, and replacing with a similar entity for both mutually exclusive and possibly coexistable events, whenever it would fit (\autoref{tab:perturbed_claim}). We then used the counterclaims to generate supporting evidence (i.e., should ideally counter the original claim) via the fine-tuned GPT-2 model. Next, we \textit{add} the planted evidence to the existing one and re-test against the original claim. 

Figure~\ref{fig:planting_supp} shows the attack's results. Contrary to~\cite{du2022synthetic}, where the accuracy always \textit{increased} after the attack, \textit{we show that it is possible to decrease it} by adding more sentences (capped at 5 after retrieval). Further,~\autoref{tab:planting_supp} shows the accuracy when removing the golden evidence and then adding the generated one. The `$\rightarrow$ NEI' ratio decreased after the addition, showing that the generated sentences can have, to some extent, the required polarity. Nonetheless, the attack has limited success, partially because counterclaims with negations could end up with evidence agreeing with the original claims, in addition to counterclaims with non-contradicting replacements~\cite{du2022synthetic} (\autoref{tab:supp_examples2}). However, in many cases, \textit{even when the generated evidence logically refutes the original claim, the model retained its predictions} (see~\autoref{tab:supp_examples} and~\autoref{tab:supp_examples3}), revealing a critical limitation we discuss next.

\begin{mdframed}[linecolor=gray(x11gray),backgroundcolor=lightgray,roundcorner=20pt,linewidth=1.5pt]
\textbf{Summary \#5:} We achieve more success in SUP to REF inversion, revealing other potential limitations. 
\end{mdframed}

\section{Discussion} \label{sec:discussion}
We here discuss the limitations and implications of our work, models' limitations, and the potential directions that we deem promising to robustify fact-verification models. 

\subsection{Limitations.} \label{sec:limitations}
\textbf{Human-in-the-loop.} We envision that fact-checking models might be more commonly used in the future as assistive solutions to fact-checkers~\cite{link_newtral} or ultimately to end-users~\cite{vo2020facts} to, e.g., output warnings. In both cases, we believe our attacks might affect humans by misleading them or denying the service. An important follow-up that we leave for future work is to evaluate the attacks by measuring such effects. This can be methodologically complex as it involves studying how to: perform these manipulations realistically and ethically, choose topics, measure the attacks' success via measuring users' perception~\cite{babaei2021analyzing}, and control for users' knowledge~\cite{pennycook2018prior} and experience (e.g., fact-checkers vs. users).

\textbf{Beyond FEVER.} FEVER allowed large-scale experiments and training. While we opted for a more comprehensive evaluation of the threat model, Du et al. show success on smaller datasets~\cite{du2022synthetic}. However, it remains unknown and ought to be evaluated how our attacks perform on other datasets with possibly different topics and characteristics.

\textbf{Wikipedia as a (Relatively) Credible Source.}
While Wikipedia can be publicly edited, it is subject to administration to remove factually wrong or biased content~\cite{link19_wiki,recasens2013linguistic}. This gives it a relative consensus of credibility compared to other online sources and makes it highly read~\cite{rosenzweig2006can}, even among fact-checkers~\cite{wineburg2019lateral}. Adversaries can exploit this wide trust to pass in disinformation or wipe traces of facts. While the Wikipedia community tirelessly resists disinformation~\cite{link11_wired}, this is not free of flaws (e.g., the Croatian Wikipedia incident~\cite{link10_wikipedia_croatia}). We hypothesize that some of our attacks (e.g., the context-preserving ones) can be stealthy even under administration. However, it can be complex to measure their potency and detection resistance  
without actual edits.

\textbf{Beyond Single-Source Datasets.} We work on Wikipedia to conform to current large-scale benchmarks; sizable datasets that match real-world fact-checking are still lacking~\cite{glockner2022missing}. In practice, fact-checkers usually rely on many sources~\cite{wineburg2019lateral}. Our attacks can, in principle, be applied to other sources~\cite{krafft2020disinformation}; however, some might not be publicly available or easy to tamper with. While our work lays the foundation for attacks in this domain, Wikipedia manipulations may affect only one of these sources, reducing the practical effect of these attacks on the whole manual fact-verification process. On the other hand, bridging this discrepancy between practitioners and automated fact-checking frameworks regarding the considered sources is one of our main takeaways that we discuss next.

\subsection{Implications}
\textbf{Ethical Considerations.} We emphasize that most of the studied attacks are based on already publicly available models, and some do not need any extra fine-tuning. Moreover, we work on a dataset containing claims that are generally not designed to be sensitive in nature, limiting any potential abuse. 

\textbf{\textit{"Only Finding Waldo"}: Models' Limitations.}
Planting attacks can considerably succeed even when as low as one evidence sentence is inserted and with the presence of the original evidence. They also succeed on instances where the model is originally highly confident. This could be partially attributed to the sparsity of golden evidence for many claims in FEVER. However, our observations on generating refuting evidence to correct claims might indicate another underlying problem. In many cases, even when the generated evidence \textit{logically refutes} the claim and the retrieved refuting evidence \textit{outnumbers} the supporting one, the model did not flip its prediction to REF (\autoref{tab:supp_examples}). At first, this can be considered a sign of robustness. However, it is possibly the exact reason why it is easy to flip the prediction of wrong claims to SUP; models might be looking for \textit{any agreement} with the evidence without considering counter stances. This is a plausible explanation, given that models were not trained with an evidence contradiction setup. However, it hints at a potential limitation in models' inference that should be investigated since \textit{the fact-checking process in practice inherently entails weighing different stances.}

\textbf{Beyond Fact-Preserving Attacks.}
While we employ attacks that target current AI vulnerabilities (e.g., imperceptible perturbations), we \textit{choose and argue for a broader scope of our work} that goes beyond adversarial examples in the sense of imperceptible perturbations and semantic equivalence. Instead, we broadly study how AI can be leveraged to create targeted disinformation and deceptive evidence manipulations at scale, impacting models and potentially humans as well.  
In such a human-centric task, simulating human-created manipulations becomes the holy grail of the attacks. These semantically driven attacks can also be more pervasive across models (see Appendix~\ref{sec:other_results}), potentially motivating their adoption by adversaries.
\textit{Even under such semantic changes, we argue that models do not show the intended behavior.} Fact-checkers do not base their verdict on a single piece of evidence, nor do they blindly trust the evidence's plausibility~\cite{shapiro2013verification}. Thus, future work should bridge this discrepancy and design `adversary-aware' defenses that better align with these practices and exploit the persisting attacks' limitations. We further explain these ideas in what follows.

\subsection{\large{How to Robustify Fact-Checking Models?}}

\textbf{Diversifying Evidence Sources.} Current camouflaging attacks need to replace the original evidence with the manipulated one; otherwise, they generally fail. Thus, in practice, fact-checking models should rely on diverse and independent sources, while also considering cross-platform coordination~\cite{donovan2019source}, to reduce the likelihood of being manipulated by a single adversarial campaign. Other evidence metadata, such as its source, should be included in the model's design~\cite{popat2018declare} to capture features such as the source's credibility, biases, or polarity. This resembles the `two-source' and `source triangulation' rules of verification in journalism~\cite{shapiro2013verification,link18_reuters}.

\textbf{Detecting Perturbations.} In addition, some camouflaging attacks can leave artifacts or perturbations enabling their identification (e.g., NLP adversarial attacks). While it might not be possible to easily recover the original evidence, human-in-the-loop systems might issue warnings about potential manipulations. The effectiveness of such warnings should also be studied~\cite{kaiser2021adapting}. On the other hand, imperceptible perturbation attacks might allow recovering the evidence upon detection via leveraging, e.g., an OCR~\cite{boucher_2022_badchars}, or utilizing recent language models that render text as images~\cite{Rust2022pixels}.

\textbf{Circular Verification against Planting Attacks.} As discussed, models should represent opposing stances among the evidence. A possible inspectable solution would be to cluster the evidence with respect to its stance~\cite{Schuster2022StretchingSN}. Moving beyond that, models should ideally contrast these opposing evidence given factors such as their source, plausibility, commonsense reasoning~\cite{jiang2021m}, and inter and intra-consistency. 

Language models may have limited factuality even in response to factual claims~\cite{lee2022factuality}. \textit{These limitations of attacks are, in fact, defense opportunities for detection based on high-level semantics}. While they all may agree with the claim, different generated samples might be inconsistent or contradicting in other details (see~\autoref{tab:supp_examples}). Similarly, a single sample might contain possibly incorrect information, beyond what supports the claim. This can fuel detection defenses of what can be called `circular fact-checking': Given the evidence, extract and verify follow-up claims and reach a plausibility decision via aggregation. This, again, echos the circular nature of information gathering and verification in investigative journalism~\cite{shapiro2013verification}, lateral reading as a fact-checkers' practice~\cite{wineburg2019lateral}, in addition to the veracity verification of manipulated media guidelines~\cite{link18_reuters}. 

\textbf{A Need for Practical Datasets.}  
As discussed in section~\ref{sec:limitations}, FEVER may be limited in matching the practical challenges of fact-verification. Therefore, there is a need to develop other datasets or augment FEVER with synthetic evidence to allow the development of models that adhere to the best practices of fact-verification. Claims should have multiple confirming or denying evidence pieces, and the evidence repositories could be partially contaminated to simulate potential adversaries. Our attacks could promisingly contribute to constructing such datasets, similar to the line of work that constructs synthetic data to help detect real fake news~\cite{huang2022faking,luo2021newsclippings}.

\textbf{Other Attacks.}
The results in section~\ref{sec:know} show that there could be a trade-off between the generated evidence complexity and the attacks' success rate. Our approach of fine-tuning and sampling can have higher success while better imitating the dataset's distribution. Other possible approaches would be to enforce the entailment between the claim and the generated evidence by training a Sequence-to-Sequence model with a verification model~\cite{bagdasaryan2022spinning}. Moreover, future work might study the effects of different prompts when generating evidence; e.g., is it possible to affect the stance of the evidence with biased variations (e.g., subtle linguistic cues~\cite{patel2021stated}) of the claim? Finally, our work is not meant to be an exhaustive evaluation of all possible attacks, as such an evaluation might be intractable and can only grow with the improvements in language generation and understanding~\cite{link_openai_ai_chatgpt}.

\section{Conclusion}
We propose a taxonomy to comprehensively study evidence and information manipulation attacks against fact-verification models. Inspired by real-life incidents of Wikipedia edits, we set the attacks' semantic \hlgoals{targets} to evidence camouflaging and planting. We then design technical methods that adversaries could utilize to achieve those targets, given the taxonomy's dimensions. Compared to previous work, we propose an extensive range of stealthier, more context-preserving, and more plausible attacks, all while simultaneously achieving higher or similar success rates and extending the attacks to all labels. We show that adversaries can decrease the performance of models even under restrictive threat models. We highlight the limitations of models' inference and discuss possible defenses by drawing insights from fact-verification in journalism. 

\begin{footnotesize}
\section*{\footnotesize{Acknowledgements}}
This work was partially funded by ELSA – European Lighthouse on Secure and Safe AI funded by the European Union under grant agreement No. 101070617. Views and opinions expressed are however those of the authors only and do not necessarily reflect those of the European Union or European Commission. Neither the European Union nor the European Commission can be held responsible for them.
\end{footnotesize}

%-------------------------------------------------------------------------------
%\bibliographystyle{plain}
%\bibliography{\jobname}
\renewcommand*{\bibfont}{\footnotesize}
\printbibliography
\appendix

%\section{Appendices}
%\setcounter{section}{0}
%\def\thesection{\Alph{section}}

\section{Implementation Details} \label{sec:impl_details}
To train the attack model $\mathcal{V}_\mathcal{A}$, we fine-tune BERT$_\text{BASE}$ or RoBERTa$_\text{BASE}$ models for 4 epochs on pairs of claims and golden evidence (for SUP and REF claims). For NEI, we pick the top 3 retrieved sentences for each claim (these should be more challenging than taking random sentences). 

To run the `lexical variation' attack, we follow the authors' \href{https://github.com/nesl/nlp_adversarial_examples}{code} and distances' hyperparameters but change the target model to RoBERTa. Words to replace are randomly sampled with probabilities proportional to the number of neighbors each word has in the embedding space~\cite{alzantot2018generating}. We adapt the `contextualized replace'\href{https://github.com/LinyangLee/BERT-Attack}{code} to the entailment task. We perturb at most 15\% of tokens in the sentence and set a probability threshold of 1.0e-5 on the BERT MLM candidates. We allow sub-words substitutes. We use an embedding distance of 0.4 between the counter-fitted vectors~\cite{mrkvsic2016counter}. For the \href{https://github.com/nickboucher/imperceptible}{`imperceptible' attacks}, we use the untargeted versions of the attack with a maximum of only 3 iterations for the genetic algorithm (vs. 10 in the original paper). Increasing the iterations' number may lead to even higher success rates; however, these attacks are expensive to run on the whole dataset ($>$ 90,000 claim-evidence pairs). For attacks against $\mathcal{V}_\mathcal{A}$, we run the attacks only on the pairs (among the top-5) where $\mathcal{V}_\mathcal{A}$'s predictions are initially correct. Since the attacks do not assume golden relevancy annotations, the labels of all retrieved sentences are set to the original claims' label (i.e., SUP or REF). We run the `imperceptible$_{\text{Ret}}$' on all the top-5 retrieved evidence to minimize the retrieval score.

For `omitting paraphrase', we use the \href{https://huggingface.co/tuner007/pegasus_paraphrase}{PEGASUS model} fine-tuned for paraphrasing. For each sentence, we generate 20 candidates using beam search (then select the lowest retrieval candidate). The \href{https://github.com/copenlu/fever-adversarial-attacks}{GPT-2 model} used in `omitting generate' and later in the `supporting generation' is trained on pairs of claims and supporting evidence for 20 epochs with a batch size of 4 and a learning rate of 0.00003. For both attacks, we use top-$k$ sampling. For `omitting generate', we also generate 20 candidates (then select the lowest retrieval candidate). For `supporting generation', we select the top 2 sentences from 160 samples (increasing the samples' number helped to have better attacks). 

For the `claim-aligned re-writing' attack, we adapt~\cite{thorne2021evidence}'s \href{https://github.com/j6mes/acl2021-factual-error-correction}{code} to re-write evidence instead of claims. We use the BERT-score masker explained in the main paper. During training, we mask the top 16 tokens. We train the T5 model for 12 epochs with a batch size of 4 and a learning rate of 0.0001. To run the attack, we mask the top 13 tokens. Depending on the masking, the T5 model re-writes single masked words or a whole span. Since this attack ideally assumes that the starting evidence is relevant, we run it only on the top 2 relevant evidence sentences. In the sampling and filtering variants, we generate 60 candidates using top-$k$ sampling. Finally, due to time and computation resources' constraints and the scale of the experimental evaluation, it is difficult to perform an exhaustive hyperparameter search. Further tuning of the hyperparameters can lead to higher success rates; our results are a lower bound.

\begin{table}[!b]
    \centering
    \vspace{-3mm}
    \resizebox{\columnwidth}{!}{%
    \begin{tabular}{l ccc|ccc|ccc} \toprule
    \textbf{Attack} & \multicolumn{3}{c}{\textbf{SUP (\%)}} & \multicolumn{3}{c}{\textbf{REF (\%)}} & \multicolumn{3}{c}{\textbf{NEI (\%)}} \\ \midrule
    &  \#1 & \#2 & \#3 & \#1 & \#2 & \#3 & \#1 & \#2 & \#3 \\ \cline{2-10}
    
    - (baseline) & 87.5 & 91.5 & 92.2 & 72.8 & 74.7 & 77.5 & 72.8 & 68.8 & 70.0 \\ \midrule
    \multicolumn{2}{l}{\textbf{Camouflaging} \inlinetiny{figs/icons_colored/goals_hide_icon} \inlinetiny{figs/icons_colored/replace_icon}} & & & & & & & &  \\ & & & & & & & & & \\
    
    Lexical variation  & 67.7 & 73.6 & 74.5 & 66.6 & 69.9 & 71.6 & - & - & - \\
    \inlinetiny{figs/icons_colored/partial_context_icon} \inlinetiny{figs/icons_colored/both_ret_ver_icon} \inlinesupertiny{figs/icons_colored/no_ext_models.pdf} & & & & & & & & & \\ & & & & & & & & & \\

    Contextualized replace & 50.0 & 55.8 & 55.6 & 60.8 & 63.9 & 64.0 &  - & - & -  \\ 
    \inlinetiny{figs/icons_colored/partial_context_icon} \inlinetiny{figs/icons_colored/both_ret_ver_icon} \inlinetiny{figs/icons_colored/ext_models_icon.pdf} & & & & & & & & & \\ & & & & & & & & & \\

    Imperceptible & & & & & & & & & \\
    \tab Homoglyph ($\epsilon=5$) & 39.9 & 60.6 & 65.1 & 52.5 & 60.1 & 60.7 & - & - & -  \\
    \tab Homoglyph ($\epsilon=12$) & 33.6 & 49.7 & 52.0 & 47.9 & 54.7 & 52.4 & - & - & -  \\ & & & & & & & & & \\
    
    \tab Reorder ($\epsilon=5$) & 37.4 & 47.7 & 49.9 & 52.3 & 54.8 & 51.4 & - & - & -  \\
    \tab Reorder ($\epsilon=12$) & 32.4 & 36.8 & 34.9 & 48.0 & 49.3 & 42.7 & - & - & -  \\ & & & & & & & & & \\

    \tab Delete ($\epsilon=5$) & 39.2 & 60.8 & 66.1 & 52.5 & 60.7 & 59.8 & - & - & -\\ 
    \inlinetiny{figs/icons_colored/full_context_icon} \inlinetiny{figs/icons_colored/both_ret_ver_icon} \inlinesupertiny{figs/icons_colored/no_ext_models.pdf} & & & & & & & & & \\ & & & & & & & & & \\

    Imperceptible$_{\text{Ret}}$ & & & & & & & & & \\
    %\tab Homoglyph ($\epsilon=5$) & 62.3 & 60.5 & - & 31.5 & 88.9 \\
    \tab Homoglyph ($\epsilon=12$) &  26.6 & 36.4 & 37.0 & 44.8 & 50.3 & 45.6 & - & - & - \\ 
    \inlinetiny{figs/icons_colored/full_context_icon} \inlinetiny{figs/icons_colored/retrieval_icon} \inlinesupertiny{figs/icons_colored/no_ext_models.pdf} & & & & & & & & & \\ & & & & & & & & & \\

    Omitting paraphrase & 50.7 & 56.8 & 55.5 & 55.8 & 60.7 & 58.4 & - & - & - \\
    \inlinetiny{figs/icons_colored/full_context_icon} \inlinetiny{figs/icons_colored/retrieval_icon} \inlinetiny{figs/icons_colored/ext_models_icon.pdf} & & & & & & & & & \\ & & & & & & & & & \\

    Omitting generate & 30.2 & 33.8 & 31.6 & 48.9 & 51.9 & 47.4 & - & - & - \\ 
    \inlinetiny{figs/icons_colored/no_context_icon} \inlinetiny{figs/icons_colored/retrieval_icon} \inlinesupertiny{figs/icons_colored/ext_models_ft_icon.pdf} & & & & & & & & & \\ \midrule
    
    \multicolumn{2}{l}{\textbf{Planting} \inlinetiny{figs/icons_colored/goals_planting} \inlinetiny{figs/icons_colored/add_icon}} & & & & & & & &  \\ & & & & & & & & & \\
    Claim-aligned re-writes & - & - & - & 36.9 & 44.9 & 42.1 & - & - & - \\ 
    %\tab +stance filtering & - & \textbf{38.4} & - & 94.4 & 1.8 \\  
    \inlinetiny{figs/icons_colored/partial_context_icon} \inlinetiny{figs/icons_colored/both_ret_ver_icon}  \inlinesupertiny{figs/icons_colored/ext_models_ft_icon.pdf} & & & & & & & & & \\ & & & & & & & & & \\

    Claim-aligned re-writes$_{\text{Ret}}$ & - & - & - & 43.1 & 48.8 & 47.6 & - & - & - \\  
    \inlinetiny{figs/icons_colored/partial_context_icon} \inlinetiny{figs/icons_colored/retrieval_icon}  \inlinesupertiny{figs/icons_colored/ext_models_ft_icon.pdf} & & & & & & & & & \\ & & & & & & & & & \\
    %& & & & & \\
    
    Supporting generation & - & - & - & 39.7 & 43.2 & 45.2 & 34.5 & 25.6 & 30.0 \\
    \inlinetiny{figs/icons_colored/no_context_icon} \inlinetiny{figs/icons_colored/verification_icon} \inlinesupertiny{figs/icons_colored/ext_models_ft_icon.pdf} & & & & & & & & & \\ & & & & & & & & & \\

    \bottomrule
    \end{tabular}}
    \caption{Attacks on CorefBERT$_\text{BASE}$ (\#1), KGAT ($\text{RoBERTa}_\text{LARGE}$) (\#2), and CorefRoBERTa$_\text{LARGE}$ (\#3).}
    \label{tab:other_models}
\end{table}

\section{Other Results and Examples} \label{sec:other_results}

In~\autoref{tab:other_models}, we show the attacks' performance (without any further adaptation) on CorefBERT$_\text{BASE}$, KGAT ($\text{RoBERTa}_\text{LARGE}$), and CorefRoBERTa$_\text{LARGE}$. Most of the attacks are still effective across models. As `imperceptible' attacks depend on the model's vocabulary, their performance can slightly degrade when transferred from BERT to RoBERTa. Increasing the perturbation budget can yield similar performance. Attacks that are based on semantically removing or adding information needed for verification are consistent across models. 

\begin{figure*}[!b]
\centering
\begin{subfigure}{0.28\textwidth}
  \centering
  % include first image
  \includegraphics[width=0.9\linewidth]{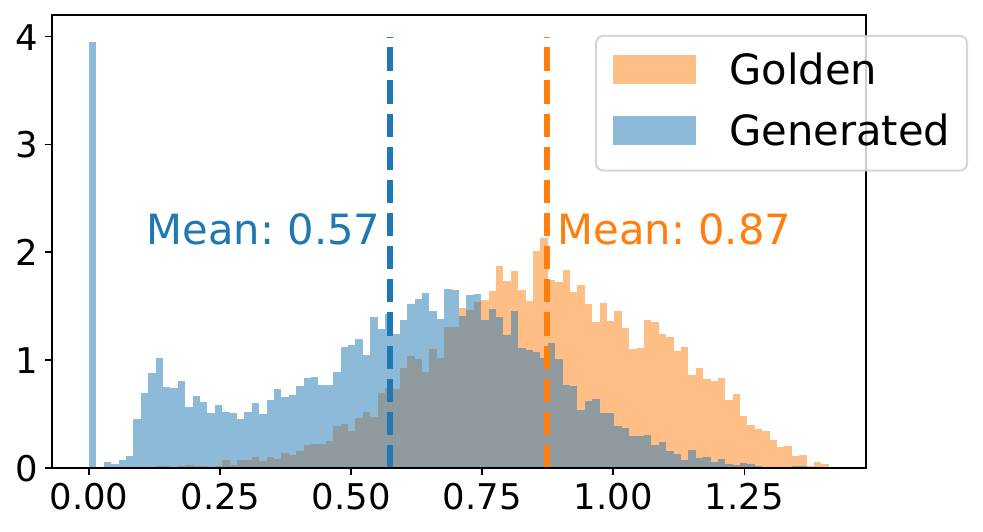} 
  \caption{Article generation~\cite{du2022synthetic}.}
\end{subfigure}
\begin{subfigure}{0.28\textwidth}
  \centering
  % include first image
  \includegraphics[width=0.98\linewidth]{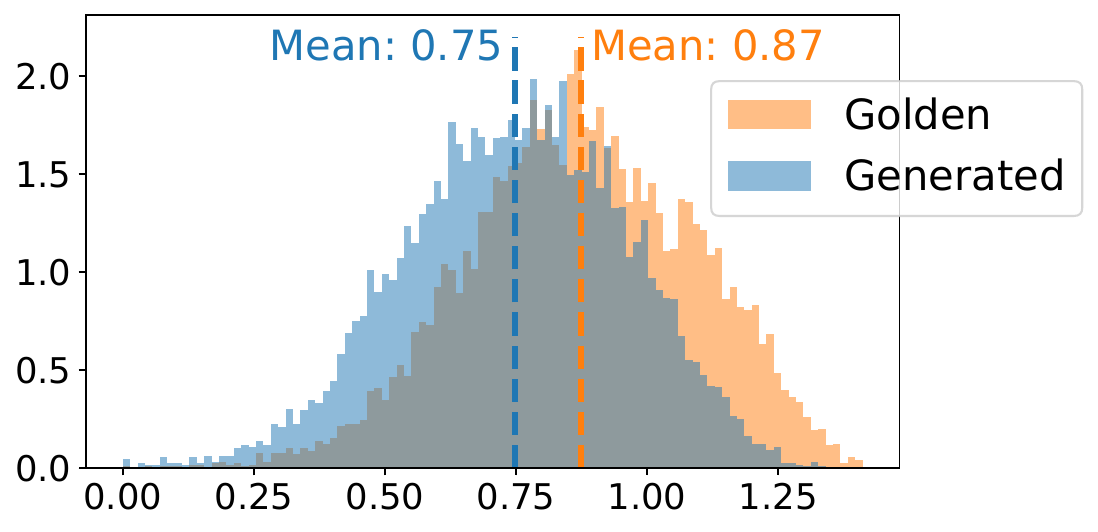}  
  \caption{Supporting generation (10\%).}
\end{subfigure}
\begin{subfigure}{0.28\textwidth}
  \centering
  % include first image
  \includegraphics[width=0.95\linewidth]{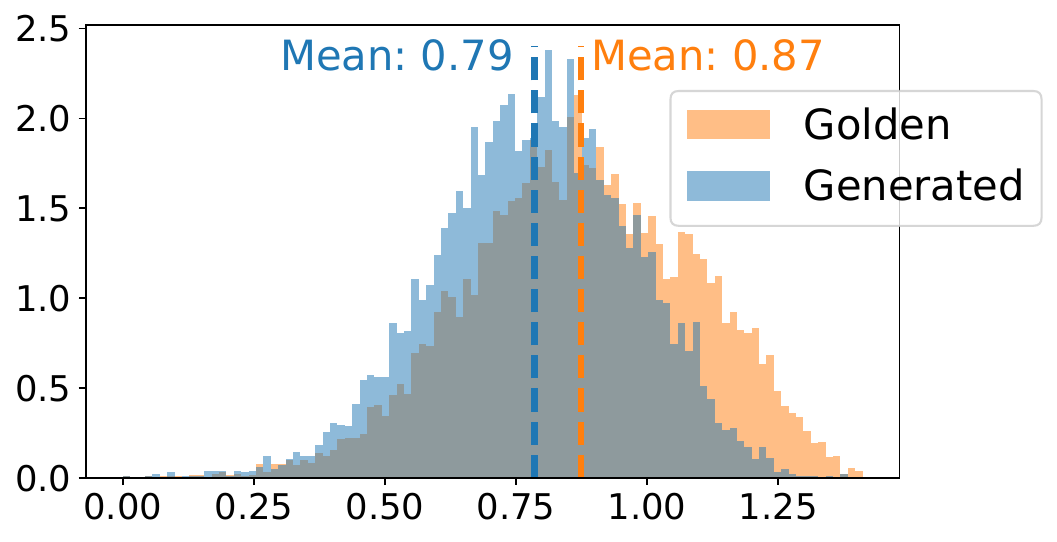}  
  \caption{Supporting generation (all).}
\end{subfigure}  

\caption{Claim-evidence embeddings' distances, in the case of generated (blue) and real-data golden evidence (orange).}
  \label{fig:embeddings}
\end{figure*}

\autoref{tab:paraphrases_example} shows examples of the automatically-created claim paraphrases (section~\ref{sec:claim_para}).~\autoref{tab:examples} shows qualitative examples (section~\ref{sec:qual}). Tables~\ref{tab:perturbed_claim},~\ref{tab:supp_examples2}, and~\ref{tab:supp_examples3} show more examples of planting attacks against the SUP label (section~\ref{sec:correct_claims}). Finally,~\autoref{fig:embeddings} shows histograms of sentence embeddings' distances between claims and evidence, for both golden and generated evidence (section~\ref{sec:know}). Our attack can lead to better matching of the golden evidence distribution compared to the baseline~\cite{du2022synthetic}. 

\newcolumntype{L}{>{\arraybackslash}m{5.5cm}}
\renewcommand{\arraystretch}{1.3}
\begin{table} [!b]
\centering
\resizebox{\linewidth}{!}{%
\begin{tabular}{L|L}
\toprule
\textbf{Original Claim} & \textbf{Paraphrase} \\ \midrule
Tilda Swinton is a vegan. & There is a person named Tilda Swinton who is a vegan. \\
Murda Beatz's real name is Marshall Mathers. & Marshall Mathers is Murda Beatz's real name. \\
Hourglass is performed by a Russian singer-songwriter. & Hourglass is a song by a Russian singer-songwriter. \\
Fox 2000 Pictures released the film Soul Food. & The film Soul Food was released by Fox 2000 Pictures. \\
Charles Manson has been proven innocent of all crimes. & Charles Manson has not been proven guilty of any crimes. \\
\bottomrule
\end{tabular}}
\caption{Automatically created claim paraphrases.} \label{tab:paraphrases_example}
\end{table}

\newcolumntype{E}{>{\arraybackslash}m{23cm}}
\begin{table*}[!t]
    \centering
    \resizebox{\textwidth}{!}{%
    \begin{tabular}{E} \toprule
    \textbf{Lexical Variation} \inlinetiny{figs/icons_colored/goals_hide_icon} \inlinetiny{figs/icons_colored/replace_icon} \inlinetiny{figs/icons_colored/partial_context_icon} \inlinetiny{figs/icons_colored/both_ret_ver_icon} \inlinesupertiny{figs/icons_colored/no_ext_models.pdf} \\ 

    \textbf{Claim:} Ann Richards was professionally involved in politics (\textbf{Label:} SUP).\\
    \textbf{Original}: Richards was the second \hlyellow{female} governor \hlyellow{of} Texas, and was frequently noted \hlyellow{in} the media \hlyellow{for} her outspoken feminism and her \hlyellow{one} liners.\newline\textbf{Edited:} Richards was the second \hlyellow{daughters} governors \hlyellow{du} Texas, and became frequently noted \hlyellow{for} the media \hlyellow{in} her outspoken feminism and her \hlyellow{eden} liners.\\ %\cdotline{1-1} 
    \midrule
    
    \textbf{Contextualized Replace} \inlinetiny{figs/icons_colored/goals_hide_icon} \inlinetiny{figs/icons_colored/replace_icon}\inlinetiny{figs/icons_colored/partial_context_icon} \inlinetiny{figs/icons_colored/both_ret_ver_icon} \inlinetiny{figs/icons_colored/ext_models_icon.pdf} \\ 
    
    \textbf{Claim:} James VI and I was a major advocate of a single parliament for Scotland and England (\textbf{Label:} SUP). \\
    \textbf{Original}: He was a \hlyellow{major} \hlyellow{advocate} of a single \hlyellow{parliament} for \hlyellow{England} and \hlyellow{Scotland}.\newline \textbf{Edited}: He was a \hlyellow{broad} \hlyellow{activist} of a single \hlyellow{legislature} for \hlyellow{Britain} and \hlyellow{Ireland}. \\ \cdotline{1-1} 

    \textbf{Claim:} Ernest Medina participated in the My Lai Massacre (\textbf{Label:} SUP).\\
    \textbf{Original}: He was the commanding officer of Company C, ... , the unit responsible for the My \hlyellow{Lai} Massacre ...\newline \textbf{Edited}: He was the commanding officer of company C, ...,  the unit responsible for the My \hlyellow{, } Massacre ...\\ \midrule 
    
    \textbf{Imperceptible/Imperceptible$_\text{ret}$} \inlinetiny{figs/icons_colored/goals_hide_icon} \inlinetiny{figs/icons_colored/replace_icon} \inlinetiny{figs/icons_colored/full_context_icon} \inlinesupertiny{figs/icons_colored/no_ext_models.pdf} \\ 
    \textbf{Claim:} Nicholas Brody is a character on Homeland (\textbf{Label:} SUP). \\
    \textbf{Edited \inlinetiny{figs/icons_colored/both_ret_ver_icon}:} Nicholas `Nick' Brody, played by actor Damian Lewis, is a fictional \hlyellow{character} on the \hlyellow{American} \hlyellow{television} series \hlyellow{Homeland} on Showtime.\\ 
    \textbf{Edited \inlinetiny{figs/icons_colored/retrieval_icon}:} \hlyellow{Nicholas} \hlyellow{`Nick'} \hlyellow{Brody}, played by actor Damian Lewis, is a fictional character on the American television series \hlyellow{Homeland} \hlyellow{on} Showtime.\\
    \midrule
    
    \textbf{Omitting Paraphrase} \inlinetiny{figs/icons_colored/goals_hide_icon} \inlinetiny{figs/icons_colored/replace_icon} \inlinetiny{figs/icons_colored/full_context_icon} \inlinetiny{figs/icons_colored/retrieval_icon} \inlinetiny{figs/icons_colored/ext_models_icon.pdf} \\ 
    
    \textbf{Claim:} Murda Beatz's real name is Marshall Mathers. (\textbf{Label:} REF). \\
    \textbf{Original}: \underline{Shane Lee Lindstrom} (born February 11, 1994) , professionally known as Murda Beatz, is a Canadian hip hop record producer from Fort Erie, Ontario. \newline
    \textbf{Edited:} Murda Beatz is a hip hop record producer from Fort Erie, Ontario. \\ \cdotline{1-1}

    \textbf{Claim:} Fox 2000 Pictures released the film Soul Food. (\textbf{Label:} SUP). \\
    \textbf{Original:} Soul Food is a 1997 American comedy drama film produced by Kenneth `Babyface' Edmonds, Tracey Edmonds and Robert Teitel and released by \underline{Fox 2000 Pictures.} \newline
    \textbf{Edited}: The 1997 American comedy drama film Soul Food was produced by Kenneth `Babyface' Edmonds, and was released by \underline{\hlred{Fox 2000 Pictures.}} \\  \midrule

    \textbf{Omitting Generate} \inlinetiny{figs/icons_colored/goals_hide_icon} \inlinetiny{figs/icons_colored/replace_icon} \inlinetiny{figs/icons_colored/no_context_icon} \inlinetiny{figs/icons_colored/retrieval_icon} \inlinesupertiny{figs/icons_colored/ext_models_ft_icon.pdf}\\
    \textbf{Claim:} Damon Albarn's debut album was released in 2011 (\textbf{Label:} REF).\\
    \textbf{Original:} Raised in Leytonstone , East London and around Colchester , Essex , Albarn attended the Stanway School , where he met Graham Coxon and eventually formed Blur , whose debut album Leisure \underline{was released in 1991} to mixed reviews.\newline
    \textbf{Edited}: Born in Leytonstone, east London, his first exposure to music came in 1991 at the age of seven, when he was discovered by Dr. Paul Barbera of St John's College in London. \\  \midrule \midrule \\
    
    \textbf{Claim-aligned Re-writing} \inlinetiny{figs/icons_colored/goals_planting} \inlinetiny{figs/icons_colored/add_icon} \inlinetiny{figs/icons_colored/partial_context_icon} \inlinetiny{figs/icons_colored/both_ret_ver_icon}  \inlinesupertiny{figs/icons_colored/ext_models_ft_icon.pdf} \\
    \textbf{Claim}: Telemundo is a English-language television network (\textbf{Label:} REF). \\
    \textbf{Original}: Telemundo is an American \underline{Spanish language} terrestrial television network owned by Comcast through the NBCUniversal division NBCUniversal Telemundo Enterprises.\newline
    \textbf{Edited}: Telemundo is an \underline{English language} television network owned by Comcast through the NBCUniversal Television Group and Comcast Enterprises. \\ \cdotline{1-1} 
    \textbf{Claim}: Juventus F.C. rejected their traditional black-and-white-striped home uniform in 1903 (\textbf{Label:} REF). \\
    \textbf{Original}: The club is the second oldest of its kind still active in the country after Genoa's football section (1893), has \underline{traditionally worn} a black and white striped home kit since 1903 and has played ... \newline
    \textbf{Edited:} The club is the second oldest of the football sections still active in the country after Genoa's football section (1893) and \underline{hasn't worn} a black and white striped home uniform since 1903 and has played ... \\ \cdotline{1-1}
    
    \textbf{Claim}: Charles Manson has been proven innocent of all crimes. (\textbf{Label:} REF). \\
    \textbf{Original}: After Manson was \underline{charged with the crimes} of which he was later \underline{convicted}, recordings of songs written and performed by him were released commercially. \newline
    \textbf{Edited:} After \underline{being proven innocent of all crimes of which he was acquitted}, recordings of songs he had performed and released were released commercially.\\ \midrule 
     
    \textbf{Supporting Generation} \inlinetiny{figs/icons_colored/goals_planting} \inlinetiny{figs/icons_colored/add_icon} \inlinetiny{figs/icons_colored/no_context_icon} \inlinetiny{figs/icons_colored/verification_icon} \inlinesupertiny{figs/icons_colored/ext_models_ft_icon.pdf}
    / \textbf{Claim-conditioned Article Generation}\cite{du2022synthetic} \inlinetiny{figs/icons_colored/goals_planting} \inlinetiny{figs/icons_colored/add_icon} \inlinetiny{figs/icons_colored/no_context_icon} \inlinesupertiny{figs/icons_colored/no_fc_icon} \inlinetiny{figs/icons_colored/ext_models_icon} \\
    
    \textbf{Claim:} Tilda Swinton is a vegan (\textbf{Label:} NEI). \\ 
    \textbf{Generated:} Swinton's work \underline{as a vegan} and as a journalist has earned her a special recognition in the media and has earned her widespread acclaim. \\ 
    \textbf{Generated}~\cite{du2022synthetic}: \underline{Tilda Swinton is a vegan.} \\ \cdotline{1-1} 
    \textbf{Claim:} Janet Leigh was incapable of writing (\textbf{Label:} REF). \\
    \textbf{Generated:} Leigh went on to study at art college in London, where she became a teacher and \underline{\hlred{writer}}. \\
    \bottomrule

    \end{tabular}}
    \caption{Samples of the attacks. `...' indicates other unchanged text. \hlyellow{Yellow highlights} are the changed words. \underline{Underlined parts} are claim-critical. \hlred{Red} indicates unsuccessful attacks according to their \hlgoals{targets}. For imperceptible attacks, we show the words where the perturbation characters were inserted.}
    \label{tab:examples}
\end{table*}

\newcolumntype{L}{>{\arraybackslash}m{6cm}}
\renewcommand{\arraystretch}{1.3}
\begin{table} [!t]
\centering
\resizebox{\linewidth}{!}{%
\begin{tabular}{L L}
\toprule
\textbf{Original Claim} & \textbf{Counterclaim} \\ \midrule
\multicolumn{2}{c}{\textbf{Mutually exclusive alternatives}} \\
Shane Black was born in 1961. & Shane Black was born in 1950. \\
The Lincoln-Douglas debates happened in Quincy, Illinois.
& The Lincoln-Douglas debates happened in Chicago, Illinois.\\ 
The Beach's director was Danny Boyle. & The Beach's director was Christopher Nolan. \\ \midrule
\multicolumn{2}{c}{\textbf{Possibly coexistable}} \\
Ann Richards was professionally involved in politics. & Ann Richards was a scientist. \\
Jack Falahee is a person who acts. & Jack Falahee is a singer. \\ \midrule
\multicolumn{2}{c}{\textbf{Opposition}} \\
James VI and I was a major advocate of a single parliament for Scotland and England. & James VI and I fought against the single parliament for Scotland and England. \\
The Cretaceous ended. & The Cretaceous still exists. \\
Ernest Medina participated in the My Lai Massacre. & Ernest Medina was against the My Lai Massacre. \\ \midrule
\multicolumn{2}{c}{\textbf{Negation}} \\
The human brain contains a hypothalamus. & The human brain does not contain a hypothalamus. \\
Rick Yune was on a tv series with Marco Polo. & Rick Yune was never on a tv series with Marco Polo. \\
\bottomrule
\end{tabular}}
\caption{Manually constructed counterclaims, used to attack SUP examples.} \label{tab:perturbed_claim}
\end{table}

\newcolumntype{D}{>{\arraybackslash}m{11.5cm}}
\begin{table}[!t]
    \centering
    \resizebox{\linewidth}{!}{%
    \begin{tabular}{D} \toprule
    \textbf{Claim}: CBS is the network that aired The Millers. \newline
    \textbf{Counterclaim}: ABC is the network that aired The Millers. \\
    \textbf{Planted \inlinetiny{figs/icons_colored/goals_planting}:} The Millers is an American period drama television series created by Peter Paige and Bradley Bredeweg which first premiered on ABC on August 4, 2002. \\ \cdotline{1-1} 
    \textbf{Claim}: Tim Roth is an English actor. \newline
    \textbf{Counterclaim}: Tim Roth is an American actor. \\
    \textbf{Planted \inlinetiny{figs/icons_colored/goals_planting}:} Timothy John Roth (born March 20, 1948) is an American actor and director. \\  \cdotline{1-1} 
    \textbf{Claim}: House is an American medical drama. \newline 
    \textbf{Counterclaim}: House is an American Sitcom. \\
    \textbf{Planted \inlinetiny{figs/icons_colored/goals_planting}:} House is an American sitcom that airs on Fox for seven seasons from May 19, 2009 to March 20, 2015. \\
    \midrule 
    \textbf{Claim}: Jack Falahee is a person who acts. \newline
    \textbf{Counterclaim}: Jack Falahee is a singer. \\
    \textbf{Planted \inlinetiny{figs/icons_colored/goals_planting}:} Jack Harold Falahee (born August 1, 1960) is an American actor and \hlred{singer}.\\ \cdotline{1-1} 
    \textbf{Claim:} In the End was positively reviewed. \newline
    \textbf{Counterclaim}: In the End was negatively reviewed. \\
    \textbf{Planted \inlinetiny{figs/icons_colored/goals_planting}:} The film received \hlred{generally positive reviews}, with \hlred{praise} directed at its performances and its ensemble cast.\\
    
    \bottomrule
    \end{tabular}}
    \caption{Counterclaims and the generated evidence. \hlred{highlighted parts} indicate attack failure.}
    \label{tab:supp_examples2}
    \vspace{-2mm}
\end{table} 

\begin{table}[!t]
    \centering
    \resizebox{\linewidth}{!}{%
    \begin{tabular}{D} \toprule
    \textbf{Claim}: House is an American medical drama. \newline 
    \textbf{Counterclaim}: House is an American Sitcom.\\ 
    
    \textbf{Original:} House (also called House , M.D.) is an American television \underline{medical drama} that originally ran on the Fox network for eight seasons, from November 16, 2004 to May 21, 2012. \\
    \textbf{Planted \inlinetiny{figs/icons_colored/goals_planting}:} House is an \underline{American television sitcom} that premiered on Fox on July 10, 2005. \\
    \textbf{Planted \inlinetiny{figs/icons_colored/goals_planting}:} House is an \underline{American television sitcom} that premiered on Fox on March 19, 1994. \\
    \textbf{Planted \inlinetiny{figs/icons_colored/goals_planting}:} House is an \underline{American sitcom} created by Matt Groening that aired on The WB Television Network from September 13, 2005 to May 19, 2009.\\
    \textbf{Original prediction}: SUP (0.99) \\
    \textbf{After-attack prediction}: SUP (0.97) \\  \midrule
    \textbf{Claim}: Black Canary is a character in comic books published by DC Comics. \newline 
    \textbf{Counterclaim}: Black Canary is a character in comic books published by Marvel. \\ 
    
    \textbf{Original:} Black Canary is a fictional superheroine in comic books published by \underline{DC Comics.} \\
    \textbf{Planted \inlinetiny{figs/icons_colored/goals_planting}:} Black Canary is a fictional superhero appearing in American comic books published by \underline{Marvel Comics.} \\
    \textbf{Planted \inlinetiny{figs/icons_colored/goals_planting}:} Black Canary (Supernatural) is a fictional character appearing in American comic books published by \underline{Marvel Comics.} \\
    \textbf{Planted \inlinetiny{figs/icons_colored/goals_planting}:} Black Canary; Canary in The A Song of Ice and Fire, by Alan Moore, is a fictional character appearing in American comic books published by \underline{Marvel Comics}, commonly in association with the superhero team the Avengers.\\
    \textbf{Original prediction}: SUP (0.98) \\
    \textbf{After-attack prediction}: SUP (0.59) \\ \midrule
    
    \textbf{Claim}: Tim Roth is an English actor. \newline 
    \textbf{Counterclaim}: Tim Roth is an American actor. \\
    \textbf{Original:} Timothy Simon Roth (born 14 May 1961) is an \underline{English actor} and director. \\
    \textbf{Planted \inlinetiny{figs/icons_colored/goals_planting}:} Timothy Francis Roth (born August 1, 1971) is an \underline{American actor}, director, and producer. \\
    \textbf{Planted \inlinetiny{figs/icons_colored/goals_planting}:} Timothy Francis `Tim' Roth (born April 9, 1941) is an \underline{American actor}, writer, producer, director, and singer.\\
    \textbf{Planted \inlinetiny{figs/icons_colored/goals_planting}:} Timothy Francis `Tim' Roth (born March 17, 1969) is an \underline{American actor}, filmmaker, and musician.\\
    \textbf{Planted \inlinetiny{figs/icons_colored/goals_planting}:} Timothy Francis `Tim' Roth (born September 9, 1967) is an \underline{American actor}, film director, screenwriter, and producer.\\
    \textbf{Original prediction}: SUP (0.96) \\
    \textbf{After-attack prediction}: SUP (0.57) \\     
    \bottomrule
    
    \end{tabular}}
    \caption{Other SUP examples where the predictions were not changed despite having retrieved refuting evidence.}
    \label{tab:supp_examples3}
\end{table} 
%%%%%%%%%%%%%%%%%%%%%%%%%%%%%%%%%%%%%%%%%%%%%%%%%%%%%%%%%%%%%%%%%%%%%%%%%%%%%%%%
\end{document}